\documentclass[prx,superscriptaddress,longbibliography,twocolumn]{revtex4-2}

\usepackage{setspace}
\usepackage{lmodern}

\usepackage[latin9]{inputenc}
\usepackage{amsmath}
\usepackage{amssymb}
\usepackage{graphicx}
\usepackage[colorlinks=True,linkcolor=blue,citecolor=blue,urlcolor=blue]{hyperref}
\usepackage{epstopdf}
\usepackage{xcolor}
\usepackage{enumerate}
\usepackage[normalem]{ulem}
\usepackage{comment}

\usepackage{longtable, booktabs} 

\begin{document}

\title{Universal bound on microwave dissipation in superconducting circuits}

\author{Thibault Charpentier}
\thanks{These two authors contributed equally}
\affiliation{Univ. Grenoble Alpes, CNRS, Grenoble INP, Institut N\'{e}el, 38000 Grenoble, France}
\affiliation{Department of Physics, University of California at Santa Barbara, Santa Barbara CA 93106, USA}
\author{Anton Khvalyuk}
\thanks{These two authors contributed equally}
\affiliation{Univ. Grenoble Alpes, CNRS, LPMMC, 38000 Grenoble, France}
\author{Lev Ioffe}
\affiliation{Google Quantum AI, Mountain View, CA}
\author{Mikhail Feigel'man}
\affiliation{Jo\v{z}ef Stefan Institute, Ljubljana 1000, Slovenia}
\affiliation{CENN Nanocenter, Ljubljana 1000, Slovenia}
\author{Nicolas Roch}
\author{Benjamin Sac\'{e}p\'{e}}
\email{benjamin.sacepe@neel.cnrs.fr}
\affiliation{Univ. Grenoble Alpes, CNRS, Grenoble INP, Institut N\'{e}el, 38000 Grenoble, France}


\begin{abstract}
\textbf{Improving the coherence of superconducting qubits is essential for advancing quantum technologies. While superconductors are theoretically perfect conductors, they consistently exhibit residual energy dissipation when driven by microwave currents, limiting coherence times. 
Here, we report an empirical scaling relation between microwave dissipation and the superfluid density, a bulk property of superconductors related to charge carrier density and disorder. Our analysis spans a wide range of superconducting materials and device geometries, from highly disordered amorphous films to ultra-clean systems with record-high quality factors, including resonators, 3D cavities, and transmon qubits.
This scaling reveals an intrinsic bulk dissipation channel, independent of surface dielectric losses, which we attribute to nonequilibrium quasiparticles trapped within disorder-induced spatial variations of the superconducting gap, with a density set by a universal material parameter. 
Our findings identify an empirical coherence limit associated with intrinsic material properties and provide a data-driven basis for materials selection in future superconducting quantum circuits.}
\end{abstract}

\maketitle

\subsection*{Introduction}

The development of scalable and fault-tolerant superconducting quantum processors hinges on the ability to maximize the coherence time of qubits, which directly impacts the fidelity and efficiency of quantum computations \cite{Devoret13,Krantz19}. Despite continuous advancements in the last two decades, the coherence time typically falls in the range of tens of microseconds to a few milliseconds~\cite{Kjaergaard2020,Ganjam24,Tuokkola2024, Dane25, Bland2025}. This limitation stems from energy dissipation mechanisms that are not yet fully understood but are believed to arise from both intrinsic and extrinsic sources in materials, circuit designs and radiations \cite{McRae2020,DeLeon21,Siddiqi2021}. 

The fundamental challenge lies in the fact that even pristine superconductors are found to lose energy when carrying an oscillating microwave current near absolute zero temperatures. These losses are characterized by the internal quality factor, $Q_{\rm i}$, the ratio between the energy stored and the energy loss per cycle of oscillation, which inversely relates to the dissipation rate. In qubits, this dissipation translates into a finite coherence time $T_1$, linked by the relation $Q_{\rm i}=\omega T_1$ where $\omega/2\pi$ is the qubit frequency.

A considerable body of work has focused on identifying and quantifying the mechanisms responsible for this residual dissipation across various superconducting materials. These mechanisms include dielectric losses~\cite{Muller19}, non-equilibrium quasiparticles~\cite{Aumentado04,deVisser2011,Barends11,Glazman21}, magnetic vortices~\cite{Song2009, Nsanzineza2014, Ku2016, Bahrami2025}, radiative losses~\cite{McRae2020,DeLeon21,Siddiqi2021}, and, more recently, mechanical vibrations~\cite{Kono2023} or substrate piezoelectricity~\cite{Ioffe04,Zhou2024}. Among these, dielectric losses --particularly those arising from defects that form a bath of two-level systems (TLS) at material surfaces and interfaces-- have garnered significant attention~\cite{Muller19}. Their ubiquitous presence and universal character in most dielectrics and oxidized surfaces~\cite{Yu88,Faoro15}, and their remarkable ability to qualitatively explain the diverse behaviors observed with increasing excitation power or temperature~\cite{Muller19}, makes them central to most analyses as the dominant source of dissipation.

However, a class of materials, such as granular aluminum (grAl) \cite{Grunhaupt18, Kristen24, Gupta2024}, titanium nitride (TiN) \cite{Coumou13,Amin2022}, amorphous indium oxide (a:InO) \cite{Charpentier2024}, or tungstene silicide (WSi) \cite{Larson2025b} known as strongly disordered superconductors \cite{Sacepe20}, defies the dielectric loss paradigm. These materials lie near a disorder-tuned superconductor-insulator transition~\cite{Sacepe20} and exhibit quality factors that decrease with disorder \cite{Charpentier2024,Gupta2024} and reach values as low as $\sim 10^3$ for the most disordered samples \cite{Charpentier2024}, only an order of magnitude higher than those reported for metallic resonators \cite{Rahim2016}. Their dissipation is found to be independent of the surrounding dielectric environment, suggesting it arises from intrinsic bulk mechanisms, sometimes referred to as inductive loss. This behavior contrasts sharply with clean elementary superconductors such as Al, Nb or Ta, which achieve $Q_{\rm i}$ values up to $10^6-10^7$ in thin-film resonators~\cite{Megrant2012, Verjauw2021, Ganjam24, Crowley23, Dane25, Bland2025}, and as high as $10^{10}$ in Nb cavities at low photon numbers and low temperatures~\cite{Romanenko20}.

Paradoxically, despite the bulk nature of dissipation in strongly disordered superconductors, their power and temperature dependencies \cite{Grunhaupt18,Amin2022,Charpentier2024} closely resemble those typically attributed to TLSs in clean materials.

This unusual bulk dissipation raises the question of whether static disorder plays a role in the intrinsic microwave dissipation in superconductors, in apparent contradiction with the standard electrodynamics theory of superconductivity \cite{Mattis58}. The latter states that a superconductor, even in the presence of strong disorder, should not dissipate energy at frequencies below twice the superconducting gap. 

In this work, we address this question through a systematic and comprehensive analysis of internal quality factors reported for conventional superconductors --including planar resonators, 3D cavities, and transmon qubits. By correlating dissipation with the inductive response, which reflects the degree of material disorder, we uncover a new fundamental limit to microwave coherence, set by bulk conductive losses driven by nonequilibrium quasiparticles.

\begin{figure*}[ht!]
\centering
 \includegraphics[width=0.81\linewidth]{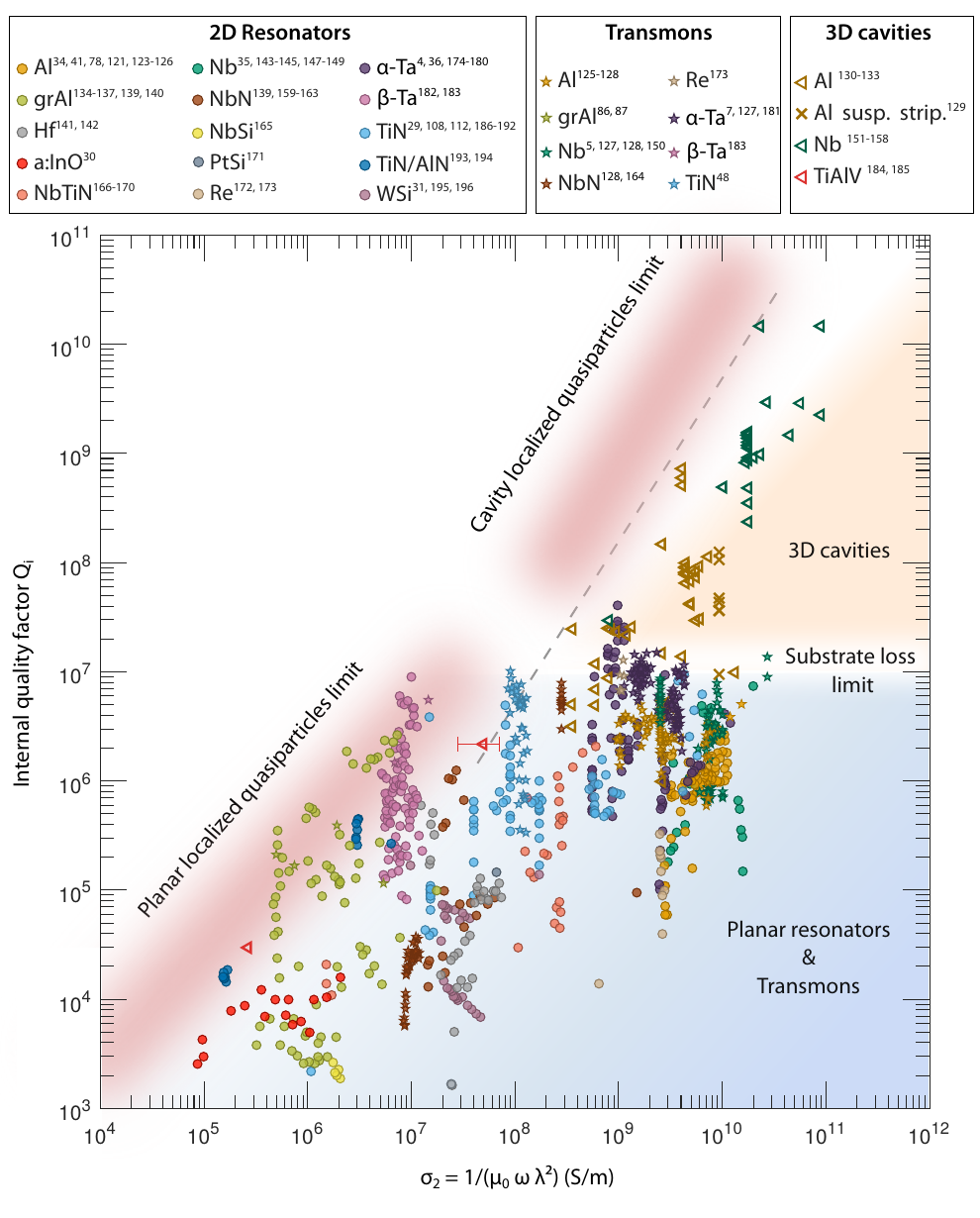}
\caption{
\textbf{Microwave dissipation from disordered to clean superconductors.} Internal quality factor \( Q_{\rm i} \) of superconducting resonators and transmon qubits measured at low temperature (\( T \lesssim 0.1 \) K) and low power, plotted as a function of the imaginary part of the complex conductivity, \( \sigma_2 = 1/\mu_0 \omega \lambda^2 \), for various superconducting materials. 
The values of the magnetic penetration depth \( \lambda \), and hence \( \sigma_2 \), span several orders of magnitude across materials, reflecting differences in disorder levels and charge carrier densities in cleaner superconductors. 
Dots represent devices in 2D geometries (coplanar waveguides or microstrip resonators), stars denote 2D transmon qubits, crosses correspond to suspended striplines, and empty triangles represent 3D resonant cavities. 
The two diagonal red stripes highlight the upper bound on microwave quality factors due to localized quasiparticles given by Eq. (\ref{eq:Qi}). For $\sigma_2 \lesssim 10^{8}$ S.m, the resulting dissipation follows the scaling $Q_{\rm i}^{\rm max} = \kappa \, \sigma_2$ with a proportionality factor $\kappa \approx (0.1-1)$ $\Omega$.m. For $\sigma_2 \gtrsim 10^{8}$ S.m, the diagonal strip reflects the scaling $Q_{\rm i}^{\rm max} \propto \sigma_2^{3/2}$, specific for small kinetic inductance fracti
on. The dashed gray line is a visual guide with slope of 3/2, fitting the best-performing cavities.
A horizontal line at \( Q_{\rm i} \approx 10^7 \) marks an apparent additional limit between planar devices (blue area) and 3D cavities (orange area), most likely related to dielectric substrate losses. Data and references are provided in Supplementary Materials Table~\ref{Table1}.
}
 \label{fig1}
\end{figure*} 


\subsection*{Superfluid density-dependent limit for microwave dissipation}
	
We begin by introducing the inductive response, a fundamental quantity that reflects how disorder influences the electrodynamics of superconductors. This response is captured by the imaginary component, $\sigma_2(\omega)$, of the complex conductivity $\sigma(\omega) = \sigma_1(\omega) - i\sigma_2(\omega)$, which directly relates to the superfluid density, $n_s$, and to the magnetic penetration depth, $\lambda$, via 
\begin{equation}
\sigma_2(\omega) = \frac{n_s e^2}{m^* \omega} = \frac{1}{\mu_0 \omega \lambda^2},
\label{eq:sigma2}
\end{equation}
where $e$ and $m^*$ are the electron charge and effective mass, respectively, and $\mu_0$ is the vacuum permeability. Physically, the oscillating supercurrent that flows over the penetration depth exhibits a kinetic inductance $L_{\rm K} = \mu_0 \lambda$, which describes the kinetic energy stored by the motion of the Cooper pair condensate. 
For thin films of thickness $d < \lambda$, the magnetic field penetrates entirely and $L_{\rm K} = \mu_0 \lambda^2 / d$.

In dirty superconductors, where the mean free path $l$ is smaller than the coherence length in the clean limit $\xi_0$, disorder reduces the superfluid density as $n_s \propto \pi n \Delta \tau/\hbar$, where $n$ is the electron density, $\Delta$ is the superconducting gap, and $\tau \ll 1/ \Delta$ is the elastic scattering time \cite{Abrikosov1959}. 
Consequently, according to Eq. (\ref{eq:sigma2}), $\sigma_2(\omega)$ decreases with increasing disorder, while the magnetic penetration depth increases \cite{Pippard1953}. This leads to an increased kinetic inductance and a weakened superfluid stiffness, which reflects a reduction in the overall superconducting phase rigidity \cite{Charpentier2024}.

Figure \ref{fig1} presents the central result of this work: the evolution of quality factors $Q_{\rm i}$ as a function of $\sigma_2$ for most of the studied materials available in the literature (see Table \ref{Table1}). The dataset includes both thin-film planar resonators (filled circles), 3D cavities (empty triangles), and transmon qubits (filled stars). To ensure consistent comparisons, $Q_{\rm i}$ values for resonators are restricted to low photon numbers and the lowest experimentally accessible temperatures (typically below $0.1$ K). This minimizes material-dependent photon-number or temperature dependencies, which, while sometimes non-monotonic, typically vary $Q_{\rm i}$ by no more than an order of magnitude. Given the multi-decade variation in both  $Q_{\rm i}$ and $\sigma_2$ on the double-logarithmic scale and the scatter in the data, accounting for these dependencies would only result in a slight upward shift in $Q_{\rm i}$ values, without altering our findings.

Two main striking observations emerge from this figure. First, the dataset is conspicuously divided by a diagonal line in the $\sigma_2 - Q_{\rm i}$ plane, above which no values of the quality factor have ever been reported. All collected values of $Q_{\rm i}$ appear to be bounded by the empirical relation $Q_{\rm i}^{\rm max} \approx \kappa \, \sigma_2$ with a proportionality factor $\kappa \sim (0.1-1)$ $\Omega$.m (see diagonal red strips), which establishes an upper limit of $Q_{\rm i}$ for a given material's $\sigma_2$. This boundary extends nearly linearly across the entire range of $\sigma_2$, that is, over seven order of magnitude, encompassing strongly disordered superconductors such as grAl (light green points), a:InO (red points), and TiN/AlN (dark blue points), moderately disordered superconductors like Hf (grey points), TiN (light blue points), and $\beta$-Ta (pink points) and even includes 3D cavities of Al (dark yellow triangles) and Nb (dark green triangles), which achieve record-high $Q_{\rm i}$ and $\sigma_2$ values. For two-dimensional films, the boundary can be expressed as 
\begin{equation}
Q_{\rm film}^{\rm max} = \frac{1}{ \eta}\frac{h}{ e^2}\sigma_2^{\square} =  \frac{1}{ \eta} \frac{h}{ e^2} \frac{1}{L_{\rm K} \omega}, 
\label{Q_film}
\end{equation}
where $\sigma_2^{\square}$ is the sheet conductance and $\eta \sim 10^{-2}$--$10^{-3}$ a dimensionless constant. The microscopic origin of $\eta$ will be discussed later in the article.

Since most superconducting resonators and cavities operate within a narrow bandwidth of $2-12$ GHz (with an average frequency of the entire dataset $\omega/2\pi = 5.5$ GHz, see Supplementary Materials Figure \ref{fig:extended_fig1}), $\sigma_2$ is primarily dependent on the material's superfluid density via the relation $\sigma_2 \propto n_s/\omega $.

Importantly, this dissipation limit is set by the best quality factors achieved for a given material or level of disorder, corresponding to a specific $\sigma_2$ value. However, a significant portion of the data lies below this boundary. These lower quality factors result from external dissipation channels that dominate and prevent the limit from being reached. These channels include standard dielectric losses, characterized by their loss tangent $\tan \delta_m$ and associated electric energy participation ratio $p_m$ (where $m$ labels the respective interface), as well as circumstantial factors such as material or sample processing imperfections. 
To quantify how these dissipation channels limit $Q_{\rm i}$, we express the quality factor as: $\frac{1}{Q_{\rm i}} = \sum_m p_m \tan \delta_m + \frac{1}{Q_{\rm MB}} + \frac{1}{Q_{\rm qp}} + \frac{1}{Q_{\rm others}}$, where $Q_{\rm MB}$ represents the Mattis-Bardeen contribution of thermal quasiparticles \cite{Mattis58}, which is typically negligible below $0.1$ K, $Q_{\rm qp}$ accounts for non-equilibrium quasiparticles, and $Q_{\rm others}$ encompasses any other sources of dissipation. 

Since dielectric losses often play a dominant role in limiting ${Q_{\rm i}}$, a common strategy to mitigate their impact consists of reducing $p_m$ by diluting electric field with large circuits. A body of work on clean superconductors has shown that reducing $p_m$ indeed enhances ${Q_{\rm i}}$ ~\cite{Wang2009, Wang15, Calusine2018, Woods19, Melville20}, up to a critical dimension beyond which ${Q_{\rm i}}$ ceases to improve and saturates \cite{Sage2011, Crowley23}. This trend is readily seen in Figure \ref{fig1} where vertically aligned clusters of data points for a given material ($\sigma_2$) illustrate this continuous increase in ${Q_{\rm i}}$ with increasing sample dimensions, corresponding to a reduction in $p_m$ (see for instance TiN, Hf, Ta). 

In stark contrast, experiments on strongly disordered films such as grAl~\cite{Grunhaupt18, Gupta2024}, a:InO~\cite{Charpentier2024, Charpentier_thesis}, TiN~\cite{Amin2022, Deng_Transmon_2023}, and WSi~\cite{Larson2025b} show no correlation between \( Q_{\rm i} \) and participation ratios, thereby ruling out dielectric loss as the dominant dissipation mechanism in the presence of disorder.

\subsection*{Dissipation from nonequilibrium quasiparticles}

The upper limit of the quality factor, observed at low photon numbers and low temperatures, scales with the material's complex conductivity --or equivalently, its superfluid density. This empirical law, which holds across both 2D circuits and 3D cavities and spans materials from strongly disordered to clean superconductors, effectively rules out dielectric loss as the dominant dissipation mechanism and provides compelling evidence for a bulk origin of dissipation --namely, conductive loss.

As a phenomenon common to all superconducting circuits, dissipation due to non-equilibrium quasiparticles emerges as a likely candidate \cite{Aumentado04,deVisser2011,Barends11,Glazman21,Mannila2021}. Various sources of such quasiparticles have been identified, including high-energy stray photons propagating through wiring  \cite{Semenov16, Catelani2019, Fischer23, Liu24,  Fischer24,Benevides24,deRooij2025a}, mechanical vibrations \cite{Yelton25} as well as cosmic radiation and radioactivity, which induce high-energy phonon down-conversion via pair-breaking \cite{Karatsu19,Vepsalainen2020,Wilen2021,McEwen2021, Cardani2021, Cardani2023, Harrington2025, Larson2025a}. Inadequate sample thermalization, a more trivial issue, could also be a contributing factor. Additionally, material imperfections and surface conditions can generate localized magnetic moments and associated sub-gap states \cite{Kharitonov12,Fominov2011}.

Since the latter are material-specific, they can be excluded as a common source. Cosmic radiation and environmental radioactivity, while universal, produces sparse quasiparticle bursts due to its slow event rate. On the other hand, most, if not all, datasets reported in Fig.~\ref{fig1} were likely obtained under similar experimental conditions, using comparable microwave components (wires, attenuators, filters, and amplifiers), many of which are now standard commercial products. This suggests that the power spectrum of unfiltered stray photons propagating through wiring and packaging is broadly similar across state-of-the-art experiments at the diagonal dissipation limit. Consequently, the quasiparticle generation rate due to pair-breaking should also be comparable.

The central question is thus the fate of these non-equilibrium quasiparticles and their contribution to dissipation. Once generated, quasiparticles rapidly relax to the superconducting gap edge, where they become trapped and recombine inefficiently. This trapping is due to sub-gap states near the gap edge, which inevitably arise in superconductors due to disorder~\cite{Larkin72,Feigelman12,Skvortsov13}. In the strong-disorder limit, these sub-gap states result from substantial spatial fluctuations of the superconducting gap~\cite{Sacepe20}, as evidenced by scanning tunneling spectroscopy on thin films of various disordered materials, including TiN~\cite{Sacepe08}, NbN~\cite{Chand12}, a:InO~\cite{Sacepe11}, and MoGe~\cite{Mandal20}, forming shallow trapping potential wells where the gap is locally suppressed~\cite{Bespalov16,DeGraaf20}. 

A recent theory predicts that the density of trapped quasiparticles scales inversely with the superconducting coherence volume, \( n_{\rm qp} \propto 1/\xi^3 \) or \( n_{\rm qp} \propto 1/(\xi^2d) \) when $\xi>d$, and only logarithmically with both the quasiparticle generation rate and the strength of electron-phonon relaxation~\cite{Bespalov16}. In realistic materials, one can reasonably assume that the inter-quasiparticle distance --beyond which localized quasiparticles do not overlap and hence fail to recombine-- cannot be much larger than $10\,\xi$ (see Supplementary Materials \ref{subsec:QP-distrib-func-and-concentation}). This implies a universal dimensionless quasiparticle density $n_{\rm qp}\xi^2 a = \eta$, with $\eta \gtrsim 0.01$--$0.001$ and $a = \min(\xi, d)$, which is in good agreement with experimental values of residual quasiparticles~\cite{Barends11,Wang2014,deVisser2014,deRooij2024}.

The microscopic contribution of trapped quasiparticles to dissipation --specifically, to the real part of the complex conductivity \( \sigma_1(\omega) \)-- remains poorly understood. 

Yet, in the case of disordered superconducting films, a simple estimate can be derived under the assumptions that localized quasiparticles are ionized into the continuum by microwave absorption. By extending the Mattis-Bardeen formula to nonequilibrium localized quasiparticles~\cite{Mattis58,Fominov2011}, one obtains a universal and constant real part of the conductivity, $\sigma_1 \sim \sigma_n n_{\rm qp} / (\nu_0 \Delta) \sim \eta  e^2/(a\hbar) $, where the conductivity $\sigma_n$ is approximated by the Drude value, $\sigma_n = e^2 \nu_0 v_F l / 3$, and the coherence length is given by $\xi^2 = v_F l \hbar /(6 \Delta) $. Here, $\nu_0$ is the density of states, and $v_F$ the Fermi velocity. This leads to: 
\begin{equation}
    Q_{\rm qp} = \frac{1}{\eta \alpha} \frac{h}{e^2} \,a\sigma_2,
    \label{eq:Qi}
\end{equation}
where \( \alpha = L_{\rm K}/(L_{\rm K}+ L_{\rm g})\) is the kinetic inductance fraction with $L_{\rm g}$ the geometric inductance. Remarkably, this result recovers the empirical trend captured by Eq. (\ref{Q_film}). Since the coherence length \( \xi \) typically varies within a factor of three across disordered materials --usually \( 5~\text{nm} \lesssim \xi \lesssim 15~\text{nm} \)-- and \( \alpha \lesssim 1 \) for devices with large kinetic inductance, this implies \( Q_{\rm qp} \sim 0.2 \, \Omega.{\rm m} \times \sigma_2 \) (with $a=\xi \sim 10$ nm and $\eta \sim 10^{-3}$). This is in excellent agreement with the observed scaling in the low-\( \sigma_2 \) regime in Fig. \ref{fig1}.

In the Supplementary Materials, we derive more accurate estimates for \( \sigma_1 \) under various dissipation mechanisms (see Supplementary Materials Fig.~\ref{fig:qp_dissip-mechanisms}) by combining realistic distribution functions for localized quasiparticles with disorder-broadened density of states~\cite{Abrikosov1960,Larkin72}. This framework provides a comprehensive classification of quasiparticle-induced losses and highlights open theoretical questions (see Supplementary Materials Fig.~\ref{fig:nonequil-qp-dissip_types_qualitative-diag}). Notably, our more elaborated estimate of the prefactor agrees well with the diagonal trend in Fig.~\ref{fig1} (see Eq.~(\ref{eq:quality-factor_numerical-estimate_dirty-limit}) in Supplementary Materials.).

Extending Eq.~\eqref{eq:Qi} to clean superconducting cavities with large \( \sigma_2 \) values (see Supplementary Materials \ref{subsec:Localized-states_direct-ionization_clean-limit}) leads to a distinct scaling, \( Q_i \propto \sigma_2^{3/2} \), due to the small kinetic inductance fraction \( \alpha \sim \omega \lambda / c \ll 1 \), which is dominated by geometric inductance (\( c \) is the speed of light). Remarkably, the best-performing cavities in Fig.~\ref{fig1} align with the corresponding slope 3/2, shown by the grey dashed line, consistent with conductive loss. 
We emphasize, nonetheless, that applying the localized quasiparticle ionization model to the clean limit (see Supplementary Materials \ref{subsec:Localized-states_direct-ionization_clean-limit}) predicts quality factors indicated by the upper red strip in Fig.~\ref{fig1}, significantly exceeding the existing data --even though it correctly captures the dependence on $\sigma_2$ related to conductive loss. Therefore, attributing the observed \( Q_i \propto \sigma_2^{3/2} \) scaling in the best-performing cavities to localized quasiparticles requires further investigation. 
A recent work~\cite{Fischer2025} indirectly confirms our conclusions by suggesting that a finite density of delocalized quasiparticles that do not recombine may also contribute to dissipation in this regime, thereby lowering the conductive-loss limit relative to that expected from localized quasiparticles alone.

An open question concerns the power dependence of resonator quality factors see, e.g., Refs.~\citep{Wang2009, hung2024}), which is commonly attributed to the saturation of TLS baths, and appears somewhat at odds with the quasiparticle-based interpretation. It has been shown that trapped quasiparticles can, in some circumstances, be described as TLS \cite{DeGraaf20}, suggesting a substantial dependence of $Q$ on the drive power (see Supplementary Material~\ref{subsec:Localized-qp_clean-limit_coherent-TLS}).
Moreover, recent experiments on grAl devices~\cite{Grunhaupt18} showed that their quality factor increases with power in a manner typically attributed to dielectric TLSs, yet exhibits no dependence on the electric field participation ratio. These observations suggest that quasiparticles, rather than dielectric TLSs, may dominate both dissipation and its power dependence --a possible reinterpretation that has received little experimental attention to date.
Ongoing theoretical work~\cite{Fominov2011,Bespalov16,Catelani2019} further supports this perspective by addressing the kinetics of non-equilibrium quasiparticles and proposing specific dissipation mechanisms, such as microwave-assisted overheating or ionization.

Another ubiquitous source of dissipation arises from vortices, which may enter superconductors during cooldown if residual magnetic fields are present. Estimating the vortex-induced dissipation, assuming optimistically low residual magnetic fields, leads to a quality factor scaling with $\sigma_2$, but with a strongly non-universal prefactor that cannot account for the observed diagonal limit (see Supplementary Materials \ref{sec:dissipation_by_vortex}).

\subsection*{Substrate loss in planar circuits}

The second striking observation that emerges from Figure \ref{fig1} is the presence of another limit specific to planar resonators: for clean superconductors with $\sigma_2 \gtrsim 10^8 \,  \text{S.m}^{-1}$, most quality factors saturate at $Q_{\rm i}\sim 10^7$, independent of their $\sigma_2$ value. This limit is highlighted by the horizontal boundary that separates the orange and blue area in Fig. \ref{fig1}. It corresponds to the saturation of quality factors upon minimizing electric-field energy in dielectric interfaces. 

The emergence of this limit suggests a common dissipation channel for planar geometry, ruling out material-dependent interface losses. Two dissipation mechanisms that are common to all microwave circuits are radiation loss and substrate loss.

As samples are most often packaged in tightly sealed (super)conductive boxes with dimensions smaller than the free-space wavelength, far-field radiation is likely suppressed, and therefore unlikely to be responsible for the limit. 

On the other hand, substrate loss in commonly used silicon and sapphire substrates has recently been thoroughly studied. For high-resistivity silicon wafers, $\tan \delta$ has been reported to be as low as $2.5\times 10^{-7}$ \cite{Woods19,Melville20}, while a recent study measured an order of magnitude higher $2.2\times 10^{-6}$ \cite{Checchin2022}. 
We emphasize that comparing silicon substrate dissipation mechanisms is challenging due to variation in wafer resistivity \cite{Zhang2024}. Moreover, the impact of intentionally introduced dopants used to homogenize wafer resistivity within the $10-20$ k$\Omega$.cm range, or other impurities (e.g. oxygen or carbon) on microwave dissipation remains poorly understood.
Thus, dissipation in silicon substrates, with these above assessments should limit the quality factor to approximately $Q_{\rm i} \sim 5 \times 10^{6}$, assuming $p_{\rm Substrate}\simeq 0.8$. This estimate is consistent with recent work that achieved $Q \sim 10^7$ on a silicon substrate \cite{Bland2025}, confirming that the exact value of loss tangent can slightly vary depending on the wafer source. For sapphire, two substrate grades obtained via different growth methods have been thoroughly studied under various processing and annealing conditions \cite{Read2023,Ganjam24}. These studies report $\tan \delta$ values ranging from $2$ to $35\times 10^{-8}$, corresponding to a substrate-limited quality factor between $3\times 10^{6}$ and $6\times 10^{7}$ \cite{Ganjam24}. 

Clearly, bulk dissipation in silicon and sapphire substrates matches the limit reached by planar superconducting microwave circuits. A recent study \cite{Ganjam24} further supports this finding, demonstrating that optimized resonator geometries, designed to minimize surface losses on the highest grade, annealed sapphire substrates, have exceeded this limit, achieving record  $Q_{\rm i}$ values up to $\sim 3 ~ 10^7$ (see Ta resonators at $\sigma_2 = 10^9$ S.m$^{-1}$ in Fig. \ref{fig1}).

\subsection*{Transmon qubits}

Interestingly, transmon qubits, which have a radically different geometry from resonators, also fall below this horizontal limit. In Fig. \ref{fig1}, we include quality factor data for transmon qubits (see filled stars), inferred from their energy relaxation times ($T_1$). The value of $\sigma_2$ is estimated from the superconductors forming the capacitor pads of the qubit. Although $\sigma_2$ is not strictly defined for transmons, we assume that the universal limit on $Q_i$ observed in resonators may also constrain transmon coherence via losses in the capacitor and wiring electrode materials --predominantly in the case of disordered, inductive superconductors, or marginally for clean superconductors-- alongside other loss mechanims. This contribution is evident in transmons made from disordered materials, as exemplified by GrAl transmons for which $Q_i\sim 10^5$~\cite{Winkel_grAl_Transmon_2020, Schon_grAl_Transmon_2020} (see filled green stars in Fig. \ref{fig1}). 

For transmons made with clean superconductors (e.g. Al, Ta, Nb, TiN), most $Q_{\rm i}$ values do not reach the diagonal limit and remain bounded at $\sim 10^7$, closely matching the upper limit observed in planar resonators (see Fig. \ref{fig1}). 
As thoroughly shown in Refs.~\cite{Deng_Transmon_2023,Bland2025}, transmon coherence time improves as $p_m$ of interfaces decreases, and can saturate at a value similar to that of resonators \cite{Deng_Transmon_2023}.
Several explanations have been proposed, including intrinsic dissipation within the Josephson junction~\cite{Deng_Transmon_2023} or losses in its surroundings due to enhanced local electric fields \cite{Ganjam24}. However, the coincidence with the $Q_{\rm i}\sim 10^7$ limit observed in resonators suggests a shared dissipation mechanism, which may rule out the Josephson junction and its dielectric environment as dominant loss sources, and instead points to bulk substrate loss as the underlying limitation. This conclusion, which emerges from Fig. \ref{fig1}, is in agreement with a recent estimate of a lower bound on the bulk-limited loss of $5.3\times 10^7$ quality factor performed on state-of-the-art transmons \cite{Bland2025}.

\subsection*{Implications for quantum circuits}

Our analysis, based on the most comprehensive dataset currently available in the literature, reveals an intrinsic upper limit to the coherence of quantum circuits for each superconducting material accessible with current technology. Because this limit scales with $\sigma_2$, the material with the largest $\sigma_2$ --niobium-- defines the highest achievable coherence once all other loss mechanisms are removed. This empirical conclusion is validated experimentally: state-of-the-art Nb 3D cavities sit at the ultimate top-right end of the diagonal boundary and reach the largest quality factors reported~\cite{Romanenko20}. Furthermore, assuming ideal resonant circuits operating at this diagonal upper limit, that is, free of dielectric losses, widely used aluminum-based technology, and even tantalum, would exhibit maximum coherence times an order of magnitude below those attainable with niobium, due to their lower $\sigma_2$.
Crucially, as our analysis shows that this limit originates from residual quasiparticles, substantial improvements in filtering~\cite{Serniak19,Vepsalainen2020,Mannila2021,Malevannaya2025}, gap engineering~\cite{Marchegiani22,pan2022,McEwen2024,Nho2025}, phonon trapping~\cite{Daddabbo_thesis,Monfardini16,Karatsu19,Henriques2019, Iaia2022, Yelton24} and packaging~\cite{Cardani2023} will be essential to push this coherence limit further.


Yet, the above arguments assume effective mitigation of interface-related dissipation and do not account for substrate dissipation, which we identify as another limiting factor for 2D planar circuits. Addressing substrate dissipation will require progress on materials quality or the development of radically new architectures, such as substrate-free suspended circuits~\cite{Pechenezhskiy2020} or on-chip 3D designs. 

At the individual qubit level, our findings indicate that coherence times on the diagonal limit could, in principle, reach $ T_1 \sim 0.1 - 1 $ s, which would represent a transformative milestone for quantum computing. However, scaling up to multi-qubit processors \cite{Mohseni2025} inevitably introduces complex couplings and additional dissipative channels. Moreover, the associated advanced fabrication techniques impose further constraints on material and interface quality. Taken together, these challenges make reaching the ultimate coherence limit extremely demanding for multi-qubit architectures.

In summary, we have identified an empirical upper bound on microwave dissipation in superconducting circuits that links the best achievable coherence to the superconducting inductive response, and thus to the material superfluid density. By compiling and analyzing results across disordered films, planar resonators, 3D cavities and transmon qubits, we show that the highest reported quality factors are constrained by a near-universal bulk conductive-loss boundary. In disordered superconductors, this behavior is consistent with nonequilibrium quasiparticles localized by disorder-induced gap inhomogeneities. In cleaner three-dimensional cavities, the best data continue to follow a conductivity-controlled trend, whereas planar devices and transmons additionally saturate near $Q_{\rm i}\sim 10^7$, consistent with substrate loss setting a separate ceiling in two-dimensional geometries. Together, these results establish a data-driven framework for identifying the intrinsic coherence ceiling of superconducting circuits and for guiding materials and architecture choices toward longer-lived quantum devices.

\bigskip

\section*{Data availability}

The code and dataset used to plot Fig. \ref{fig1} are available at Zenodo 10.5281/zenodo.15737011.

\bigskip

\section*{Acknowledgments}

We thank D. Basko, A. Dreuzet, Q. Ficheux, R. Gao, L. Glazman, M. Houzet, J.Meyer, I. Pop, and M. Scheffler for valuable discussion. We thank R. Gao, N. de Leon, and I. Pop for sharing their data. B.S. thanks Y. Chen for many valuable discussions.
B.S. and T.C. has received funding from the European Union's Horizon 2020 research and innovation program under the ERC grant SUPERGRAPH No. 866365. N.R. has received funding from the European Union's Horizon 2020 research and innovation program under the ERC grant SuperProtected No. 101001310. N.R. and B.S. acknowledge funding from the ANR agency under the 'France 2030' plan, with Reference No. ANR-22-PETQ-0003. A.V.K. is grateful for the support by Laboratoire d'excellence LANEF in Grenoble (ANR-10-LABX-51-01).

\section*{Competing Interests} The authors declare that they have no competing interests.

\bibliography{Biblio}

\bigskip
\clearpage

\section*{Supplementary Materials}


\subsection{Complex conductivity}

Here, we introduce the key quantities that describe the electrodynamics of superconductors. In the London gauge, the London equation $\mathbf{j_s} = \omega \sigma_2 \mathbf{A}$ relates the superconducting current density $\mathbf{j_s}$ to the vector potential $\mathbf{A}$ where the proportionality constant $\sigma_2$ is the imaginary part of the complex conductivity $\sigma(\omega)$. 
 A direct consequence of this equation is that magnetic fields can enter a superconductor only over the magnetic penetration depth $\lambda = 1 / \sqrt{ \mu_0 \omega \sigma_2}$. The surface impedance of a superconductor $Z_s = 1/ \lambda \sigma(\omega)$ consists mainly in a dissipationless inductive channel, with characteristic inductance $L_K = 1/\omega \lambda \sigma_2 = \mu_0 \lambda$, called the kinetic inductance, that describes the kinetic energy stored by the Cooper pair condensate. For thin films of thickness $d < \lambda$, the magnetic field penetrates entirely and $L_K = 1/\omega d \sigma_2 = \mu_0 \lambda^2 / d$.
 
 
\subsection{Microwave loss due to nonequilibrium quasiparticles in clean and
moderately disordered superconductors}

In this section, we describe the possible loss mechanisms arising
from nonequilibrium quasiparticles, with a focus on distinguishing
between localized and delocalized quasiparticles. The specific dissipation
mechanism associated with these quasiparticles depends on a combination
of material properties, experimental conditions, and the characteristics
of the external quasiparticle source. These factors conspire to establish
a steady state of the quasiparticle subsystem. Remarkably, the most
relevant regimes can be classified based on a small set of physical
parameters.

Subsections~\ref{subsec:Quasiparticle-DoS}-\ref{subsec:Loc-deloc_recombination_limit}
introduce the key features of the quasiparticle density of states
(DoS) and distribution function, respectively. Subsection~\ref{subsec:Qualitative-picture}
provides a qualitative overview of dissipation mechanisms, classifying
them by their relative dissipation strength (see also Supplementary
Materials Fig.~\ref{fig:nonequil-qp-dissip_types_qualitative-diag}).
The subsequent subsections, \ref{subsec:Localized-states_direct-ionization_dirty-limit}--\ref{subsec:Localize-qps_low-freq_possible-loss-channels},
examine each mechanism in detail and, where possible, provide estimates
for the associated quality factor. We also briefly review theoretical
models describing the power dependence of the quality factor, as this
behavior further accentuates the qualitative differences between dissipation
mechanisms.

Although we have aimed to provide a comprehensive overview of the
relevant physical mechanisms, the crude nature of some of our estimates
--or the absence of estimates in certain regimes-- combined with
the empirical trends observed in Fig.~\ref{fig1}, should serve as
a strong motivation for further theoretical and experimental investigations
into the kinetics of localized quasiparticles.

Our analysis does not include systems with substantial concentration
of magnetic impurities, both in the bulk and on the surface. The primary
reason is that the best-performing devices are often obtained by taking
active measures to suppress additional channels of dissipation. In
particular, they are usually characterized by negligible magnetic
disorder~(see Ref.~\citep{DeGraaf20} and references therein).
In the outlook subsec.~\ref{subsec:Localize-qps_low-freq_possible-loss-channels}, we briefly return to the subject of bulk magnetic impurities.

We also emphasize that strongly disordered superconductors, such as
high-resistance a:InO films, are not described by the semiclassical
picture of superconductivity~\citep{feigelmanFractalSuperconductivityLocalization2010,Sacepe20,Charpentier2024}
and are thus outside the scope of this section. In these films, the dissipation may be dominated by disorder-induced localized collective excitations~\citep{Khvalyuk24,khvalyuk2025}.

\bigskip{}

\subsubsection{Quasiparticle Density of States}\label{subsec:Quasiparticle-DoS}

One of the key quantities for understanding quasiparticle-induced
dissipation is the quasiparticle DoS per spin projection $\nu(E)$.
It is well known~\citep{DeGraaf20,Larkin72,Skvortsov13} that $\nu(E)$
in a conventional superconductor--- either clean ($l\gg\xi=\hbar v_{F}/(\pi\Delta_{0})$,
where $v_{F}$ is the Fermi velocity) or moderately dirty ($\xi=\sqrt{D\hbar/(2\Delta_{0})}\gg l\gg\lambda_{F}$,
where $D=v_{F}l/3$ is the diffusion coefficient, and $\lambda_{F}$
is the Fermi wave length) consists of two parts (see Supplementary
Material Fig.~\ref{fig:qp_dissip-mechanisms}\textbf{a}):

\emph{i)}~A sharp BCS peak, $\nu_{\text{bulk}}(E)$, associated with
the spectrum of \emph{delocalized} quasiparticle states. This part
of the DoS ends at the mobility edge $E_{g}$, which
lies slightly below the bulk superconducting order parameter $\Delta_{0}$.
Both the difference $\varepsilon_{g}=\Delta_{0}-E_{g}$ and the finite
broadening of the BCS peak originate from nonmagnetic impurity scattering~\citep{Larkin72}.

\emph{ii)}~A rapidly decaying tail of \emph{localized} states $\nu_{\text{loc}}(E<E_{g})$,
arising from rare spatial regions where disorder suppresses the superconducting
order parameter locally, forming shallow potential wells with bound
states that can trap quasiparticles. The characteristic spatial extent
$r_{\text{loc}}(E)$ of such states and their corresponding DoS $\nu_{\text{loc}}(E)$
are given, for the clean ($\xi\ll l$) and dirty ($\xi\gg l$) limits
respectively~\citep{DeGraaf20,Skvortsov13,Bespalov16,Fischer2025},
by:

\begin{widetext}
	\begin{equation}
		\nu^{\text{clean}}_{\text{loc}}(\varepsilon)=\nu_{T}de^{-\left(\varepsilon/\varepsilon_{T}\right)^{5/4}}\,\,\,(d\ll\xi),\,\,\,\frac{r^{\text{clean}}_{\text{loc}}(\varepsilon)}{2\xi}\sim\left(\frac{\Delta_{0}}{\varepsilon}\right)^{1/2},\label{eq:localization-properties_clean-limit}
	\end{equation}
	\begin{equation}
		\nu^{\text{dirty}}_{\text{loc}}(\varepsilon)=\begin{cases}
			\nu_{T}d\left(\varepsilon/\varepsilon_{T}\right)^{1/2}e^{-\left(\varepsilon/\varepsilon_{T}\right)^{5/4}}, & d\ll\xi,\\
			\nu_{T}\left(\varepsilon/\varepsilon_{T}\right)^{9/8}e^{-\left(\varepsilon/\varepsilon_{T}\right)^{3/2}}, & d\gg\xi,
		\end{cases}\,\,\,\frac{r^{\text{dirty}}_{\text{loc}}(\varepsilon)}{2\xi}=\left(\frac{2\Delta}{3\varepsilon}\right)^{1/4},\label{eq:localization-properties_dirty-limit}
	\end{equation}
\end{widetext}

\noindent  $\varepsilon=E_{g}-E\ll\Delta_{0}$ is the distance
to the mobility threshold, and $\nu_{T},\varepsilon_{T}$ are the
characteristic magnitude and energetic extent of the tail that depend
on the material properties. Eq.~\eqref{eq:localization-properties_clean-limit}
is applicable to thin films ($d\ll\xi$), while Eq.~\eqref{eq:localization-properties_dirty-limit}
features expressions both for thin films and for bulk superconductors~($d\gg\xi$).

A different form of the stretched exponential DoS tail in the dirty
limit~\citep{meyerGapFluctuationsInhomogeneous2001,Skvortsov13}
may be realized in the normal part of the SNS junctions~\citep[Sec. IVC2]{Skvortsov13}.
However, this regime is never realized in thin superconducting films
and bulk superconductors~\citep[Sec. IVC3]{Skvortsov13}. Moreover,
this does not affect the applicability of our analysis, so we will
use the Eq.~\eqref{eq:localization-properties_dirty-limit} for all
estimations.

Associated with the localized part of the spectrum is the maximum
concentration of localized quasiparticles that a given material can
host in a steady state: 
\begin{equation}
	n^{\max}_{\text{loc}}=\intop^{E_{g}}_{0}dE\,\nu_{\text{loc}}(E).\label{eq:loc-tail-capacity}
\end{equation}
Note that the usual factor of 2 accounting for spin degeneracy is
absent here. This is because two quasiparticle states occupying the
same localized state would recombine rapidly, allowing only single
occupancy of such a state.

\subsubsection{Quasiparticle distribution function and concentration\label{subsec:QP-distrib-func-and-concentation}}
Another important physical quantity is the steady-state quasiparticle
concentration: 
\begin{equation}
	n_{\text{qp}}:=2\intop^{+\infty}_{0}dE\,\nu(E)\,f_{s}(E),\label{eq:qp-concentration-def}
\end{equation}
where $f_{s}(E)$ is the steady-state distribution function of the
quasiparticles. We emphasize that both $f_{s}$ and, consequently,
$n_{\text{qp}}$, are difficult to describe theoretically, as they
depend critically on the nature of the occupied states due to the
complexity of the underlying kinetics.

It is reasonable to assume that \emph{delocalized} quasiparticles
rapidly equilibrate with phonons. In this case, $f_{s}$ is approximately
given by a quasi-equilibrium distribution function (assuming $T_{\text{eff}}\ll\Delta_{0}$):
\begin{equation}
f_{s}(E>E_{g})\approx\frac{x_{\text{qp}}}{1+e^{E/T_{\text{eff}}}}\approx x_{\text{qp}}e^{-E/T_{\text{eff}}},\label{eq:deloc-qps_quasi-equilib_distrib-function}
\end{equation}
where the prefactor $x_{\text{qp}}$, determining the concentration
of delocalized quasiparticles, strongly depends on both material properties
(such as phonon-assisted recombination and inelastic relaxation rates)
and on the nature of the external quasiparticle source (e.g., high-energy
phonons from environmental radiation). Moreover, the effective temperature
$T_{\text{eff}}$ may differ significantly from the base temperature
$T_{0}$, due to insufficient thermal coupling between the phonon
subsystem and the external cooling bath.

The situation is drastically different in the case where exclusively
localized states are occupied. Due to the exponentially decaying spatial
profile of the localized wavefunction, the recombination rate $\Gamma_{r}$
of such quasiparticles (e.g., via phonon emission) is strongly suppressed.
Indirect experimental evidence suggests values of $\Gamma_{r}$ as
low as $\sim0.1\,\text{MHz}$ in \emph{clean} NbN films~\citep{DeGraaf20}.
This implies that even a weak external source can maintain a stationary
concentration of localized quasiparticles that is many orders of magnitude
higher than the thermal equilibrium concentration at the same temperature.

A qualitative model describing this scenario was proposed in Ref.~\citep{Bespalov16}
for the dirty limit.
The key observation is that the recombination rate of a localized
quasiparticle due to any local process (e.g., phonon emission) can
be estimated as $\Gamma_{r}(E)\sim\bar{\Gamma}\,n_{\text{qp}}\,r^{3}_{\text{loc}}(E),$
where $\bar{\Gamma}$ is the normal-state electron-phonon relaxation
rate. This estimate follows from the requirement of finite spatial
overlap between the wavefunctions of recombining quasiparticles. Due
to the strong energy dependence of $r_{\text{loc}}(E)$, see Eqs.~\eqref{eq:localization-properties_clean-limit}--\eqref{eq:localization-properties_dirty-limit},
the recombination rate $\Gamma_{r}(E)$ becomes much smaller than
the single-particle energy relaxation rate as the quasiparticle relaxes
to lower energies.

As a result, Ref.~\citep{Bespalov16} demonstrated that quasiparticle
relaxation effectively halts at an energy $E_{C}<E_{g}$ (see Supplementary
Materials Fig. \ref{fig:qp_dissip-mechanisms}\textbf{a}). Quasiparticles with
energy above $E_{C}$ either relax to $E_{C}$ or recombine before
reaching it, while recombination at $E<E_{C}$ is negligible due to
the vanishing overlap between localized states. Further relaxation
to $E<E_{C}$ is also blocked: the quasiparticle occupies the ground
state of its local well, and phonon-assisted tunneling to deeper wells
with lower energy is prevented, as those states are already occupied.
Recombination remains the only possible relaxation mechanism if the
quasiparticles have opposite spins.

This picture is summarized by the following ansatz for the steady-state
distribution function: 
\begin{equation}
f_{s}(E)\approx\frac{1}{2}\theta(E_{C}-E).\label{eq:localized-qps_steady-state_distrib-func}
\end{equation}
The prefactor $1/2$ reflects the fact that each localized state can
host only one quasiparticle: the presence of two quasiparticles in
the same state would lead to significant spatial overlap and rapid
recombination, on a timescale much shorter than $\Gamma^{-1}_{r}$.
Equation~\eqref{eq:localized-qps_steady-state_distrib-func} implies
a vanishing effective temperature $T_{\mathrm{eff}}$, although poor
phonon thermalization may still result in a finite $T_{\mathrm{eff}}$
due to a residual balance between phonon heating and recombination
processes for $E>E_{C}$. A more accurate expression for $f_{s}(E)$
would require a detailed microscopic model of localized quasiparticle
kinetics, which lies beyond the scope of this work.

The cutoff energy $E_{C}$ is determined by the balance between external
quasiparticle generation and recombination power. However, because
the steady-state quasiparticle concentration $n_{\mathrm{qp}}$, defined
in Eq.~\eqref{eq:qp-concentration-def}, is a steep function of $E_{C}$
{[}owing to the exponential tail of $\nu_{\text{loc}}(E)$, see Eqs.~\eqref{eq:localization-properties_clean-limit}--\eqref{eq:localization-properties_dirty-limit}{]},
the value of $E_{C}$ is only weakly sensitive to microscopic parameters.
Instead, it is essentially fixed by the condition that the localized
wavefunctions do not spatially overlap at energy $E\approx E_{C}$:
\begin{align}
	& V_{\text{loc}}(E_{C})\,n_{\text{qp}}\approx V_{\text{loc}}(E_{C})\intop^{E_{C}}_{0}dE\,\nu_{\text{loc}}(E)\sim1,\label{eq:stationary-concentration_main-equation}\\
	& V_{\text{loc}}(E)=r^{2}_{\text{loc}}(E)\,\min\left\{ d,r_{\text{loc}}(E)\right\} ,\label{eq:loc-volume_general-expr}
\end{align}
where $V_{\mathrm{loc}}(E)$ represents the localization volume and
accommodates both thin films (finite $d$) and bulk materials ($d\to\infty$).
The precise numerical factor on the right-hand side depends on microscopic
details such as drive strength, shape of localized states, and recombination
mechanisms, but these affect $E_{C}$ only logarithmically due to
the steep energy dependence of $\nu_{\mathrm{loc}}(E)$. This observation
renders the ansatz in Eq.~\eqref{eq:localized-qps_steady-state_distrib-func}
self-consistent.

The crucial practical consequence of Eqs.~\eqref{eq:localization-properties_clean-limit}--\eqref{eq:localization-properties_dirty-limit},
and~\eqref{eq:stationary-concentration_main-equation}--\eqref{eq:loc-volume_general-expr}
is that localized quasiparticles are most likely found in bound states
of characteristic size $r_{\mathrm{loc}}(E_{C})$ at a certain energy
$E_{C}$. This leads to a universal steady-state concentration:
\begin{equation}
	r_{\text{loc}}(E_{C})=b\xi\Leftrightarrow n_{\text{qp}} \sim n^{\text{uni}}_{\text{loc}} = \frac{1}{b^{2}\xi^{2}\min\left\{ d,b\xi\right\} },\label{eq:stationary-concentration_main-eq_2}
\end{equation}
where $b=3-10$ is the characteristic size of the localized state
at energy $E_{C}$, expressed in units of the coherence length $\xi_{0}$.
The value of $b$ depends only weakly on the strength of the external
drive or on other microscopic details~\citep{Bespalov16}.

For two-dimensional films of thickness $d\ll\xi$, the value of
$b$ acquires a logarithmic correction $b^{3}\to b^{2}+2\ln d/\xi$
in the dirty limit~\citep[Supp. Note B]{Fischer2025} due to a different
spatial profile of the optimal fluctuation providing the finite DoS~\eqref{eq:localization-properties_dirty-limit}.
Moreover, the applicability of the effective long-wave-length theory
leading to Eq.~\eqref{eq:localization-properties_dirty-limit} for
$d\ll\xi$ is marginal in 2D~\citep[Sec. IVB]{Skvortsov13}, so further
logarithmic corrections can arise. However, for the estimates of
the present work, all these effects can be neglecting, so Eq.~\eqref{eq:stationary-concentration_main-eq_2}
will be used throughout.

\subsubsection{Recombination of localized and delocalized quasiparticles can limit
	the maximum concentration\label{subsec:Loc-deloc_recombination_limit}}

Importantly, Ref.~\citep{Bespalov16} describes the equilibrium between
the production and recombination rates of localized quasiparticles.
However, the recently published Ref.~\citep{Fischer2025} identifies
recombination between localized and \emph{delocalized} quasiparticles
as an important ingredient in the formation of the corresponding steady state.
Because known quasiparticle sources (see the main text) initially populate
delocalized states, nonequilibrium quasiparticles are expected to relax
into the localized DoS tail only at a later stage. The main insight of
Ref.~\citep{Fischer2025} is that recombination between these temporarily
\emph{delocalized} quasiparticles and the localized ones sets an upper
limit on $n_{\text{qp}}$ in the steady state, even when the latter is
dominated by localized quasiparticles:
\begin{equation}
	n_{\text{qp}}\le n^{\text{dyn}}_{\text{qp}}\sim2\nu_{0}\Delta_{0}\,\frac{\Gamma_{\text{relax}}}{\Gamma_{\text{loc-deloc}}},
\end{equation}
where $\Gamma_{\text{relax}}$ and $\Gamma_{\text{loc-deloc}}$ are the
relaxation and localized-delocalized recombination rates, respectively.
When both rates are governed by phonon-mediated processes,
$n^{\text{dyn}}_{\text{qp}}$ can be estimated as~\citep{Fischer2025}:
\begin{equation}
	n^{\text{dyn}}_{\text{qp}}\sim2\nu_{0}\Delta_{0}\,\left(\varepsilon_{T}/\Delta_{0}\right)^{3/2+\mu}\label{eq:dynamic-limit_estimate}.
\end{equation}
Here, the exponent $\mu$ depends on whether the typical phonon momentum
transfer~$q$ satisfies $ql\gg1$ or $ql\leq1$, where $l$ is the electron
mean free path. Specifically, $\mu=2$ for sufficiently clean superconductors
with $ql\gg1$, whereas for more disordered superconductors with $ql\leq1$,
one finds $\mu=3$ if the impurities move with the lattice, and $\mu=1$
if the impurities are effectively decoupled from it; see
Ref.~\citep[Supp. Note A]{Fischer2025} and references therein for further details.

For relatively clean materials, $n^{\text{dyn}}_{\text{qp}}$ provides
a more stringent upper bound on $n_{\text{qp}}$ than $n^{\max}_{\text{loc}}$.
Using Eqs.~\eqref{eq:localization-properties_dirty-limit}--\eqref{eq:loc-tail-capacity},
together with the expression for $\nu_{T}$ in the dirty case~\citep{Larkin72,Bespalov16},
$n^{\max}_{\text{loc}}$ can be estimated as
\begin{equation}
	n^{\max}_{\text{loc}}\sim2\nu_{0}\Delta_{0}\,\frac{\left(\varepsilon_{T}/\Delta_{0}\right)^{3/2}}{\varepsilon_{g}/\Delta_{0}},
\end{equation}
where $\varepsilon_{g}=\Delta_{0}-E_{g}>0$. For relatively clean
materials, one has $\mu>0$ and $\varepsilon_{g}/\Delta_{0}\ll1$,
which implies $n^{\max}_{\text{loc}}\gg n^{\text{dyn}}_{\text{qp}}$.

More importantly, the dynamic upper bound $n^{\text{dyn}}_{\text{qp}}$
can be incompatible with the universal value~\eqref{eq:stationary-concentration_main-eq_2}.
Using the estimates of Ref.~\citep{Fischer2025} for aluminum, one
obtains $n^{\text{dyn}}_{\text{qp}}\xi^{3}\sim3\times10^{-7}$. This
value is at least four orders of magnitude smaller than that given by
Eq.~\eqref{eq:stationary-concentration_main-eq_2}, thereby ruling out the
universal behavior in the clean materials on the right-hand side of
Fig.~\ref{fig1} of the main text as a steady state \emph{under a constant
quasiparticle source}. This is consistent with the observed dissipation
strength; see Subsec.~\ref{subsec:Dissipation_deloc-qps}.

Crucially, $n^{\text{dyn}}_{\text{qp}}$ is expected to grow rapidly
with disorder and may therefore exceed the universal limit~\eqref{eq:stationary-concentration_main-eq_2}
in more disordered films. Indeed, using the results of Ref.~\citep{Skvortsov13},
the value of $\varepsilon_{T}$ can be roughly estimated as
\begin{equation}
\varepsilon_{T}/\Delta_{0}\sim\begin{cases}
g^{-8/5}_{\xi}, & d\gg\xi,\\
g^{-4/3}_{\xi}, & d\ll\xi,
\end{cases}
\end{equation}
where $g_{\xi}=\hbar\sigma_{0}\max\left\{ \xi,d\right\} /e^{2}$ is
the dimensionless conductance. Note that $\varepsilon_{T}/\Delta_{0}$
increases sharply with disorder, as measured by the inverse dimensionless
conductance $g^{-1}_{\xi}$. Consequently, the ratio of $n^{\text{dyn}}_{\text{qp}}$,
Eq.~\eqref{eq:dynamic-limit_estimate}, to the universal limit~\eqref{eq:stationary-concentration_main-eq_2}
also grows rapidly with disorder:
\begin{equation}
\frac{n^{\text{dyn}}_{\text{qp}}}{n^{\text{uni}}_{\text{loc}}}\sim\begin{cases}
b^{3}g^{-7/5-8/5\mu}_{\xi}, & d\gg\xi,\\
b^{2}g^{-1-4/3\mu}_{\xi}, & d\ll\xi.
\end{cases}
\end{equation}
As a result, in more disordered films one expects
$n^{\text{dyn}}_{\text{qp}}/n^{\text{uni}}_{\text{loc}}\gg1$, making the
universal limit achievable. Nevertheless, a reliable conclusion would
require an experimental determination of the actual value of $\varepsilon_{T}$.
While some estimates of $\varepsilon_{T}$ exist~\citep{gurra2025,Fischer2025}
for clean Al and Nb systems, the authors are not aware of any reliable
probe of $\varepsilon_{T}$ in more disordered systems.

In contrast to the universal limit~\eqref{eq:stationary-concentration_main-eq_2},
$n^{\text{dyn}}_{\text{qp}}$ is much more sensitive to the specific
relaxation and generation mechanisms. In particular, the approach of
Ref.~\citep{Fischer2025} and, consequently, the limiting value $n^{\text{dyn}}_{\text{qp}}$,
are not applicable to burst-like sources of high-energy quasiparticles
(e.g. cosmic rays in high-quality qubits~\citep{DeGraaf20}): after
a single high-energy event, delocalized quasiparticles are rapidly
eliminated from the system through recombination and relaxation, and
soon disappear in the absence of an external source. The remaining localized
quasiparticles obey the slow-recombination picture of Ref.~\citep{Bespalov16},
exhibiting the universal concentration~\eqref{eq:stationary-concentration_main-eq_2},
with only a slow logarithmic time dependence~\citep{DeGraaf20} (provided
the localized tail can accommodate such a concentration, i.e., $n^{\text{uni}}_{\text{loc}}<n^{\max}_{\text{loc}}$).
This further emphasizes the still incomplete theoretical understanding of localized quasiparticle kinetics.

\subsubsection{Qualitative picture of quasiparticle dissipation\label{subsec:Qualitative-picture}}

Using five parameters ---$T_{\text{eff}},\varepsilon_{T},n^{\max}_{\text{loc}},n^{\text{dyn}}_{\text{qp}},n_{\text{qp}}$
and the external frequency $\omega$--- Supplementary Materials Fig.~\ref{fig:nonequil-qp-dissip_types_qualitative-diag}
provides a classification of the main dissipation mechanisms associated
with nonequilibrium quasiparticles. The primary classification criterion
is whether the delocalized part of the quasiparticle spectrum is significantly
occupied. This can occur in two scenarios: \emph{i)}~$n_{\text{qp}}>n^{\max}_{\text{loc}}$,
implying that localized states are insufficient to accommodate all
quasiparticles, and therefore delocalized states must be populated;
or \emph{ii)}~$T_{\text{eff}}>\varepsilon_{T}$, where the effective
temperature is sufficient to ionize localized states and thus create
a significant concentration of delocalized quasiparticles.

In either case, the resulting dissipation is captured by a Mattis-Bardeen-type
expression~\citep{Mattis58,Fominov2011}, 
\begin{equation}
\frac{\sigma_{1}}{\sigma_{0}}\sim\frac{n_{\text{qp}}}{\nu_{0}\omega}\;\;\Leftrightarrow\;\;\sigma_{1}\sim\frac{\Delta_{0}}{\omega}\,n_{\text{qp}}\xi^{3}_{0}\,\frac{2e^{2}}{\hbar\xi_{0}},\label{eq:sigma-1_Mattis-Bardeen}
\end{equation}
describing absorption by transitions in the continuous spectrum, as
detailed in subsec.~\ref{subsec:Dissipation_deloc-qps}.

In the localized regime, the dissipation mechanism becomes strongly
frequency-dependent, as illustrated in Supplementary Materials Fig.~\ref{fig:nonequil-qp-dissip_types_qualitative-diag}.
However, it seems likely that in many relevant devices the driving
frequency $\omega$ is large enough to ionize localized states directly,
i.e., $\omega>E_{g}-E_{C}\sim\varepsilon_{T}$. As a result, the dissipation
is again described by Eq.~\eqref{eq:sigma-1_Mattis-Bardeen}, but
with a universal quasiparticle density (see Eq.~\eqref{eq:stationary-concentration_main-eq_2}).

An important note concerns the relation between $\sigma_{1}$ and
the quality factor $Q$, given by:
\begin{equation}
Q=\frac{1}{\alpha}\frac{\sigma_{2}}{\sigma_{1}},\label{eq:quality-factor_with-alpha}
\end{equation}
where $\alpha$ is the kinetic inductance fraction $\alpha=L_{K}/(L_{K}+L_{g})$,
and $L_{g}$ is the geometric inductance~\citep{Zmuidzinas12}. The value of $\alpha$ depends
on the device geometry and material parameters, with $\alpha\approx1$
for dirty superconductors and $\alpha\sim2\omega\lambda/c\ll1$ in
the clean limit, where $\lambda$ is the London penetration depth.
Nonetheless, some dirty superconductors with $l\ll\xi_{0}$ may still
have a large enough superfluid density to yield small $\alpha$, as
in the TiN samples of Ref.~\citep{Richardson2020} (see Supplementary
Materials Table~I).

As the frequency is lowered below $\omega<\varepsilon_{T}$, direct
ionization is no longer allowed, resulting in a rapid increase in
the quality factor. Residual dissipation in this regime may arise
from other mechanisms illustrated in Supplementary Materials Figs.~\ref{fig:qp_dissip-mechanisms}\textbf{d-f}
and~\ref{fig:nonequil-qp-dissip_types_qualitative-diag}. While these
mechanisms remain poorly understood, they are unlikely to yield $\sigma_{1}$
values larger than those associated with direct ionization, as we
show below.

A notable example is the excitation of localized quasiparticles between
discrete bound states within a single potential well --effectively
forming a two-level system (TLS)-- as proposed in Ref.~\citep{DeGraaf20}
for clean superconductors (see Supplementary Materials Fig.~\ref{fig:qp_dissip-mechanisms}\textbf{d}).
In subsec.~\ref{subsec:Localized-qp_clean-limit_coherent-TLS}, we
estimate the corresponding quality factor, which turns out to be much
larger than that associated with direct ionization.

Other speculative loss channels are briefly discussed in subsec.~\ref{subsec:Localize-qps_low-freq_possible-loss-channels},
see also Supplementary Materials Fig.~\ref{fig:qp_dissip-mechanisms}\textbf{d-f}.

In addition to estimating the quality factors corresponding to these
scenarios (as done below), a direct experimental probe of $f_{s}(E)$,
$n_{\text{qp}}$ would help identify the correct case. However, such
probes for localized quasiparticles remain elusive.

Most experimental probes of quasiparticle density are performed in
clean devices in the right part of Fig.~\ref{fig1}
of the main text, which, as we argue in subsec.~\ref{subsec:Dissipation_deloc-qps},
are dominated by the delocalized quasiparticles. These probes amount
to extracting the quasiparticle concentration from the dissipation
data~\citep{Barends11,deVisser2011,Wang2014,deVisser2014,connolly2024},
thus implicitly relying on their delocalized character (in subsec.~\ref{subsec:Dissipation_deloc-qps}
we perform similar estimates). A recent work~\citep{gurra2025} attempted
extracting $\nu_{\text{loc}}(E)$ from the (S/N)IS tunneling junction
$I(V)$ curves of clean Al and Nb devices. Importantly, the result suffers
from inaccuracies due to parasitic processes, such as phonon-assisted
tunneling, so the extracted localized DoS tail only serves as a loose upper bound on the true $\nu_{\text{loc}}(E)$. Nevertheless, using
this estimate, Ref.~\citep{gurra2025} also concludes that the observed
dissipation does not originate from localized quasiparticles; therefore,
no information can be extracted about $f_{s}(E)$ in the localized
regime.

In contrast, the experimental method of Ref.~\citep{DeGraaf20} relies
on characterizing individual two-level TLS. By assuming that these
TLS originate from localized quasiparticles, this method gives indirect
insights into both $\nu_{\text{loc}}(E)$ and $f_{s}(E)$. The results
appear consistent with their theoretical interpretation.

\subsubsection{Direct ionization in the dirty limit\label{subsec:Localized-states_direct-ionization_dirty-limit}}

Following Ref.~\citep[Ch. 9 and 11]{Kamenev_book}, we briefly recall
the key elements relevant in the dirty limit, $l\ll\xi$. The superconducting
state for a given spatial profile of the order parameter $\Delta(r)$
is described by the Usadel equation for the retarded Green's function
$g^{R}(\omega,r)$, a matrix in Nambu space. The diagonal component,
$\nu_{0}g^{R}_{11}$, corresponds to $\langle\psi\bar{\psi}\rangle$,
while the off-diagonal component, $\nu_{0}g^{R}_{12}$, corresponds
to $\langle\psi\psi\rangle$. The local DoS is given by $\nu(\omega,r)=\nu_{0}\mathrm{Re}\,g^{R}_{11}(\omega,r)$.
The Green's function satisfies the nonlinear normalization condition
$g^{2}=1$, which can be conveniently parametrized using the complex
spectral angle $\theta(\omega,r)$: namely, $g^{R}_{11}=\cos\theta$,
and $g^{R}_{12}=\sin\theta$. The Usadel equation is solved self-consistently
along with the gap equation for $\Delta$.

The presence of short-range disorder generates an effective depairing
term in the long-wavelength limit of the Usadel equation~\citep{Larkin72},
and induces uncorrelated noise in the order parameter superimposed
on its uniform background value. For energies in the quasiparticle
continuum, these fluctuations can be neglected, allowing one to consider
a spatially uniform Green's function. In this case, the Usadel equation
reduces to: 
\begin{equation}
\cos\theta+i\sin\theta\,\frac{E+i0}{\Delta}=\frac{\eta}{2}\sin2\theta,\label{eq:GF_AG_mean-field}
\end{equation}
where $\eta$ is the dimensionless depairing parameter. The resulting
DoS, $\nu_{\text{AG}}(E)=\nu_{0}\,\mathrm{Re}\,\cos\theta(E)$, follows
the Abrikosov-Gor'kov profile: a smeared BCS peak with a sharp gap
edge located at $E_{g}\approx\Delta\left(1-\eta^{2/3}\right)^{3/2}$~\citep{Abrikosov1960,Larkin72}.

At energies below the gap, $E<E_{g}$, a finite DoS arises only due
to \emph{rare} local dips in the order parameter. The most probable
configuration leading to a subgap state at energy $E=E_{g}-\varepsilon$,
with $\varepsilon\ll E_{g}$, is found via the method of optimal fluctuations~\citep{Larkin72,Skvortsov13},
and has the form:
\begin{equation}
	\theta(\varepsilon,r)=\frac{\pi}{2}+\sqrt{\frac{\varepsilon}{E_{g}}\,}\phi\left(\frac{1}{1-\eta^{2/3}}\,\frac{r}{r_{\text{loc}}(\varepsilon)}\right),\label{eq:GF_LO_instanton}
\end{equation}
where $\phi\sim1$ is a real-valued function that depends on the dimensionality
of the system, and $r_{\text{loc}}(\varepsilon)$ is defined in Eq.~\eqref{eq:localization-properties_dirty-limit}.
Taking into account the statistical weight of such fluctuations leads
to the subgap DoS $\nu_{\text{loc}}(E)$ given by Eq.~\eqref{eq:localization-properties_dirty-limit}
in the dirty limit.

The real part of the conductivity due to transitions into the quasiparticle
continuum can be obtained from the Kubo formula~\citep{Kamenev_book,Fominov2011}
using the nonequilibrium Keldysh Green's function:
\begin{widetext}
\begin{equation}
	\frac{\sigma_{1}(\omega)}{\sigma_{0}}=-\frac{1}{8\omega}\intop dE\,\text{Im}\text{Tr}\left\{ \tau_{3}g^{R}(E)\tau_{3}g^{K}(E-\omega)+\tau_{3}g^{K}(E)\tau_{3}g^{A}(E-\omega)\right\} ,\label{eq:nonequilibrium-conductivity_Kubo}
\end{equation}
\end{widetext}

\noindent where $g^{K}=g^{R}\,F-F\,g^{A}$, with $F=F_{L}\tau_{0}+F_{T}\tau_{3}$,
and $\tau_{i}$ are Pauli matrices in Nambu space. The functions $F_{L}$
and $F_{T}$ are the longitudinal (odd in energy) and transverse (even
in energy) components of the distribution function, respectively.
Both vanish in thermal equilibrium at zero temperature.

In the nonequilibrium case due to quasiparticle generation, the total
quasiparticle concentration remains small, so we neglect the back-action
of these quasiparticles on $g^{R,A}$, using the equilibrium solution
of the Usadel equation. We further assume that the transverse component
$F_{T}$, associated with particle-hole imbalance~\citep{Schmid-Schon,Kamenev_book},
relaxes quickly and can be set to zero. Thus, the nonequilibrium effect
enters solely through $F_{L}$, which becomes: 
\begin{equation}
	F_{L}=\tanh\left\{ E/2T\right\} -2f_{s}(E)\overset{T\to0}{\longrightarrow}1-2f_{s}(E),
\end{equation}
where $f_{s}(E)$ is the steady-state quasiparticle distribution function.
Substituting this into Eq.~\eqref{eq:nonequilibrium-conductivity_Kubo},
one arrives at the simplified expression derived in Eq.~(47) of Ref.~\citep{Fominov2011},
now applied to a \emph{nonequilibrium} quasiparticle distribution
(with $E_{\pm}=E\pm\omega/2$):
\begin{widetext}
	\begin{equation}
		\frac{\sigma_{1}(\omega)}{\sigma_{0}}=\frac{1}{\omega}\intop^{+\infty}_{0}dE\,\left[F_{L}(E_{+})-F_{L}(E_{-})\right]\,\left[\text{Re}g^{R}_{11}(E_{+})\text{Re}g^{R}_{11}(E_{-})+\text{Im}g^{R}_{12}(E_{+})\text{Im}g^{R}_{12}(E_{-})\right].\label{eq:nonequilibrium-conductivity_Kubo_simplified}
	\end{equation}
\end{widetext}

As explained above Eq.~\eqref{eq:localized-qps_steady-state_distrib-func},
an external agent generates quasiparticles that eventually relax into
localized states at energies $E\sim E_{C}$, resulting in a distribution
function approximated by $f_{s}(E)\approx\frac{1}{2}\theta(E_{C}-|E|)$.
Owing to this fact, and to the steep energy dependence of the localized
density of states, the integrand in Eq.~\eqref{eq:nonequilibrium-conductivity_Kubo_simplified}
is sharply peaked near $E=E_{C}$. To evaluate the integral, we average
over fluctuations of the order parameter: $g^{R}(E_{-})$ is approximated
using the optimal fluctuation form from Eq.~\eqref{eq:GF_LO_instanton},
while $g^{R}(E_{+})$ is taken in the uniform (mean-field) approximation
given by Eq.~\eqref{eq:GF_AG_mean-field}. This is justified because
the dominant fluctuations of $\Delta(\boldsymbol{r})$ that contribute
to the DoS at $E\sim E_{C}$ do not significantly modify the continuum of
states at $E>E_{g}$. Substituting into Eq.~\eqref{eq:nonequilibrium-conductivity_Kubo_simplified}
and performing the averaging yields:
\begin{widetext} 
\begin{equation}
\frac{\sigma_{1}(\omega)}{\sigma_{0}}=\frac{2}{\omega}\int^{\infty}_{0}dE\,\left[f_{s}(E_{-})-f_{s}(E_{+})\right]\frac{\nu_{\text{loc}}(E_{-})}{\nu_{0}}\frac{\nu_{\text{AG}}(E_{+})}{\nu_{0}}\sim\frac{1}{\omega}\int^{E_{C}}_{0}dE\,\frac{\nu_{\text{loc}}(E)}{\nu_{0}}\frac{\nu_{\text{AG}}(E+\omega)}{\nu_{0}},\label{eq:sigma-1_MT-like-answer_diry-case}
\end{equation}
\end{widetext}

\noindent The last term is a smooth function of energy, so the integral
can be estimated using Eq.~\eqref{eq:stationary-concentration_main-equation},
yielding: 
\begin{equation}
\frac{\sigma_{1}(\omega)}{\sigma_{0}}\sim\frac{\nu_{\text{AG}}(E_{C}+\omega)/\nu_{0}}{\nu_{0}\omega\,r^{2}_{\text{loc}}(E_{C})\,\min\left\{ d,r_{\text{loc}}(E_{C})\right\} }.\label{eq:sigma-1_estimation_dirty-limit}
\end{equation}

The frequency dependence of this result arises from the shape of the
smeared BCS peak $\nu_{\text{AG}}$, which governs the availability
of final states for the ionization process. The corresponding quality
factor is then estimated using Eq.~\eqref{eq:quality-factor_with-alpha}:
\begin{align}
	Q^{\text{dirty}}_{\text{ioniz}} & \sim\frac{b^{2}}{\alpha}\frac{\nu_{0}\omega}{\nu_{\text{AG}}T_{c}}\frac{\hbar\sigma_{2}}{e^{2}}\min\left\{ d,b\xi\right\} ,\label{eq:quality-factor_estimation_dirty-limit}\\
	& \sim\left(0.05-2\right)\times10^{-4}\,\frac{f\left[\text{GHz}\right]}{\alpha\,T_{c}\left[\text{K}\right]}\,\sigma_{2}\left[\text{S/m}\right]\nonumber \\
	& \times\min\left\{ \frac{d}{5},\xi\right\} \left[\text{nm}\right],\label{eq:quality-factor_numerical-estimate_dirty-limit}
\end{align}
Here, we restore $\hbar$ and use $\hbar D/\xi^{2}\sim\Delta_{0}\sim T_{c}$,
$\sigma_{0}=2\nu_{0}e^{2}D$, and $r_{\text{loc}}\approx b\xi$. For
the numerical estimate, Eq.~\eqref{eq:quality-factor_numerical-estimate_dirty-limit},
we take $\nu_{0}$ as the typical value of $\nu_{\text{AG}}$ away
from the spectral edge~\citep{Larkin72}, and assume $b=3-10$, which
accounts for the given range in the prefactor. We use $b=5$ within
the minimum function to approximate the thickness dependence.

The exact ratio $\nu_{0}/\nu_{\text{AG}}$ depends not only on $\omega$
and $E_{C}$, but also on the precise shape of the nonequilibrium
distribution $\delta f$, which causes additional smearing of $\nu_{\text{AG}}$.
Equivalently, one may interpret this as an uncertainty in the definition
of $E_{C}$ that is comparable to the BCS peak width. This leads to
an effective $\nu_{0}/\nu_{\text{AG}}\sim1$.

To contextualize Eq.~\eqref{eq:quality-factor_numerical-estimate_dirty-limit}
in terms of Fig.~\ref{fig1} of the main text, we consider
TiN devices (see Supplementary Materials Table~I or Ref.~\citep{Gao2022}),
which yield $Q^{\text{dirty}}_{\text{TiN}}\sim(0.03-1.2)\times10^{6}$
for $\xi_{0}\approx20\,\text{nm}$. The upper end of this range aligns
well with the experimental value for this film. This indicates that
the proposed mechanism can account for a significant fraction of the
dissipation observed in state-of-the-art dirty superconducting devices.

Unfortunately, the value of $Q$ on its own cannot be used to determine
the type of the occupied state (to this end, an independent probe
of $\nu_{\text{loc}}(E)$, $f_{s}(E)$ is required). However, the delocalized
regime with total $n_{\text{qp}}$ significantly smaller than the universal limit $n^{\text{uni}}_{\text{loc}}$, Eq.~\eqref{eq:stationary-concentration_main-eq_2}, is only possible in the presence of a mechanism that excites localized
quasiparticles (e.g., a nonzero~$T_{\text{eff}}$~\citep{Fischer2025},
see also Supplementary Material Fig.~\ref{fig:nonequil-qp-dissip_types_qualitative-diag}).
Such a mechanism is then likely to generate quaisparticles on its
own (as it the case for finite $T_{\text{eff}}$~\citep[Supp. Note A]{Fischer2025}),
rendering the total~$n_{\text{qp}}$ significantly larger than $n^{\text{uni}}_{\text{loc}}$
and thus delivering $Q$ much lower than Eq.~\eqref{eq:quality-factor_numerical-estimate_dirty-limit}
(see subsec.~\ref{subsec:Dissipation_deloc-qps}). Moreover, explaining
the quality factors of the best-performing devices in Fig.~\ref{fig1} of the main text across a variety of materials by a \emph{universal}
(in units of $\xi^{-3}$) concentration of delocalized quasiparticles
would require a very special coincidence in the nonuniversal factors
providing finite $n_{\text{qp}}$ in each particular device.

Nevertheless, the uncertainties in the distribution function $f_{s}(E)$
and in the value of $b$ in Eq~\eqref{eq:quality-factor_estimation_dirty-limit},
---both arising from the poorly understood kinetics of localized
quasiparticles--- necessitate further investigation.

\subsubsection{Direct ionization in the clean limit when $\alpha\ll1$\label{subsec:Localized-states_direct-ionization_clean-limit}}

We expect that Eqs.~\eqref{eq:sigma-1_estimation_dirty-limit}--\eqref{eq:quality-factor_numerical-estimate_dirty-limit}
remain qualitatively valid in the clean limit, $\xi\ll l$, as well.
In this regime, the derivation of $\sigma_{1}$ should be reformulated
in terms of the direction-resolved Green's function $g(E,\boldsymbol{n})$,
where $\boldsymbol{n}$ is a unit vector on the Fermi surface, and
the corresponding Eilenberger equation for the retarded/advanced components~\citep{DeGraaf20}.
Averaging over disorder-induced fluctuations in the order parameter
$\Delta(r)$, which are responsible for the existence of localized
states, is expected to yield a result similar to Eq.~\eqref{eq:sigma-1_MT-like-answer_diry-case},
with all subsequent approximations and transformations remaining applicable.

For comparison with experimental data in Fig.~\ref{fig1} of the
main text, we must also account for the additional dependence of $Q$
on $\omega$ and $\sigma_{2}$ via the kinetic inductance fraction
$\alpha\ll1$. For a cubic 3D cavity resonator with side length $L=\pi c/\omega$
corresponding to the fundamental mode, the kinetic inductance fraction
can be estimated as: 
\begin{equation}
	\frac{1}{\alpha}=\frac{L}{6\lambda}\approx0.18\sqrt{\frac{\sigma_{2}\left[\text{S/m}\right]}{f\left[\text{GHz}\right]}},\label{eq:alpha-dependence_numeric}
\end{equation}

In the clean limit, the film thickness satisfies $d\gg\xi$, so
Eq.~\eqref{eq:quality-factor_numerical-estimate_dirty-limit} reduces
to: 
\begin{equation}
Q^{\text{clean}}_{\text{ioniz}}\sim2\times10^{-5}\,\sigma^{3/2}_{2}[\mathrm{S/m}]\cdot\xi_{0}[\mathrm{nm}].\label{eq:Q-estimation_clean-limit_final-estimate}
\end{equation}
In this estimate, we have dropped explicit dependence on $\omega$
and $T_{c}$, as their influence is expected to be smaller than the
overall uncertainty introduced by the chain of approximations. Consequently,
it is difficult to assign a reliable error range to Eq.~\eqref{eq:Q-estimation_clean-limit_final-estimate}.

Nonetheless, this formula predicts quality factors that are at least
two orders of magnitude higher than those measured in aluminum cavities
with $\sigma_{2}=4\times10^{9}\,\mathrm{S/m}$ and $\xi=100\,\mathrm{nm}$,
and the discrepancy increases with $\sigma_{2}$, since the dependence
is faster than linear. While the numerical disagreement at specific
values of $\sigma_{2}$ may be attributed to the approximate nature
of Eq.~\eqref{eq:Q-estimation_clean-limit_final-estimate}, the broader
trend suggests that dissipation in these systems is dominated by a
stronger mechanism --such as delocalized quasiparticles, discussed
in subsec.~\ref{subsec:Dissipation_deloc-qps} below.

\subsubsection{Coherent TLS Rabi oscillations in the clean limit\label{subsec:Localized-qp_clean-limit_coherent-TLS}}

For low frequencies in clean systems (see Supplementary Materials
Figs.~\ref{fig:qp_dissip-mechanisms}\textbf{d}~and~\ref{fig:nonequil-qp-dissip_types_qualitative-diag}),
the analysis of Subsec.~\ref{subsec:Localized-states_direct-ionization_dirty-limit}
must be modified. Instead of using the Kubo formula for direct ionization,
Eq.~\eqref{eq:nonequilibrium-conductivity_Kubo}, one must calculate
$\sigma_{1}$ arising from an ensemble of \emph{coherent} two-level
systems (TLSs), formed by the first two bound quasiparticle states
in a typical local minimum of the order parameter~\citep{DeGraaf20}.

The power absorbed by a single TLS with energy splitting $\omega_{0}$,
dephasing rate $\Gamma_{2}$, and relaxation rate $\Gamma_{1}$, is
described by a Lorentzian profile~\citep[Sec. 2.3]{Shnirman}:
\begin{equation}
P=\frac{1}{2}\left(\boldsymbol{p},\boldsymbol{E}\right)^{2}\,\omega\,\tanh\left\{ \frac{\omega_{0}}{2T}\right\} \,\frac{\Gamma_{2}}{\left(\omega_{0}-\omega\right)^{2}+\Gamma^{2}_{2}},\label{eq:single-dipole_absorbed-power}
\end{equation}
where $\boldsymbol{p}$ is the transition dipole moment and $\boldsymbol{E}$
is the applied electric field. At zero temperature ($T=0$), an ensemble
of such TLSs with randomly oriented dipole moments contributes to
the dissipative conductivity as
\begin{equation}
\sigma^{\text{TLS}}_{1}(\omega)=\intop d\omega_{0}\,n(\omega_{0})\frac{\bar{p}^{2}(\omega_{0})}{3}\frac{\omega\Gamma_{2}/2}{\left(\omega_{0}-\omega\right)^{2}+\Gamma^{2}_{2}},
\end{equation}
where $\bar{p}(\omega_{0})$ is the typical dipole moment of a TLS
with splitting $\omega_{0}$, and $n(\omega_{0})$ is the spectral
density of such TLSs. The latter is determined by the number of localized
quasiparticles occupying the ground state of local order parameter
dips with excitation energy $\omega_{0}=\zeta\varepsilon$, and is
given by:
\begin{equation}
n(\omega_{0})=\frac{1}{\zeta}\left.f_{s}(E)\,\nu_{\text{loc}}(E)\right|_{\varepsilon\equiv E_{g}-E=\omega_{0}/\zeta},\label{eq:clean-case_TLS_spectral-density}
\end{equation}
where $\zeta\approx0.7$~\citep{DeGraaf20} is the ratio between
the first excitation energy and the ground state energy in a typical
optimal well. As before, we assume the steady-state distribution $f_{s}(E)\approx\theta(E_{C}-E)/2$,
cf.~Eq.~\eqref{eq:localized-qps_steady-state_distrib-func}.

The transition dipole moment for such a TLS is estimated as
\begin{equation}
\bar{p}(\omega_{0})\sim\left.e\frac{\varepsilon}{\Delta_{0}}\,r_{\text{loc}}(E)\right|_{\varepsilon=\omega_{0}/\zeta},
\end{equation}
where the prefactor $\varepsilon/\Delta_{0}$ accounts for the near
equal mixing of electron and hole components for quasiparticles near
the gap edge.

Since $\Gamma_{2}\ll\varepsilon_{T}$ (the typical energy scale of
the localized DoS~\citep{DeGraaf20}), the Lorentzian is sharply
peaked and the integral simplifies:
\begin{equation}
\sigma^{\text{TLS}}_{1}(\omega)\sim\frac{\pi\omega}{6\zeta}f_{s}(E)\nu_{\text{loc}}(E)\left[e\frac{\varepsilon}{\Delta_{0}}r_{\text{loc}}(E)\right]^{2},
\end{equation}
with $\varepsilon=E_{g}-E=\omega/\zeta$.

We emphasize that this estimate is considerably less reliable than
Eq.~\eqref{eq:sigma-1_estimation_dirty-limit} for the dirty limit.
The frequency dependence is sensitive to the detailed shape of both
$f_{s}(E)$ and $\nu_{\text{loc}}(E)$, and anisotropic features of
the optimal fluctuations in the clean limit~\citep{DeGraaf20} are
not accounted for --neither in the kinetic treatment nor in the transition
matrix elements.

We now employ the specific profile~\eqref{eq:localization-properties_clean-limit}
of $\nu_{\text{loc}}$ in Eq.~\eqref{eq:stationary-concentration_main-equation}
to express the integral in terms of $\nu_{\text{loc}}(E_{C})$:
\begin{equation}
\varepsilon_{T}\,V_{\text{loc}}(E_{C})\,\nu_{\text{loc}}(E_{C})\,\left(\varepsilon_{C}/\varepsilon_{T}\right)^{-1/4}\sim1,
\end{equation}
which allows us to estimate the quality factor at $\omega\sim\zeta(E_{g}-E_{C})$,
where $\sigma_{1}$ is \emph{maximal}:
\begin{equation}
Q^{\text{clean}}_{\text{TLS}}\sim\left(3-10\right)\,\frac{1}{\alpha}\left(\frac{\Delta_{0}}{\varepsilon_{C}}\right)^{2}\frac{\hbar\sigma_{2}\xi}{e^{2}}.\label{eq:Q-estimation_clean-limit}
\end{equation}
Here, we use that $\varepsilon_{C}=E_{g}-E_{C}\sim\varepsilon_{T}$
within our precision, and take $b=3-10$ as before. We also assume
the condition $d\gg b\xi$ holds, which is typical for 3D resonators.
However, sufficiently thin Al films may violate this assumption (see
Supplementary Materials Table~I).

In analogy to Eq.~\eqref{eq:Q-estimation_clean-limit_final-estimate},
the dependence of $\alpha$ in the clean limit {[}cf.~Eq.~\eqref{eq:alpha-dependence_numeric}{]}
should be included, leading to:
\begin{equation}
Q^{\text{clean}}_{\text{TLS}}\sim\left(0.2-3\right)\,10^{-3}\,\frac{\sigma^{3/2}_{2}\left[\text{S/m}\right]\xi\left[\text{nm}\right]}{\sqrt{f\left[\text{GHz}\right]}}.\label{eq:clean-case_TLS_quality-factor}
\end{equation}
In this estimate, we used $\varepsilon_{C}/\Delta_{0}=0.05-0.1$ as
the characteristic depth of the relevant fluctuations of the order parameter.
Both this uncertainty and that in $b$ contribute to an overall uncertainty
of about an order of magnitude.

It is instructive to compare Eq.~\eqref{eq:clean-case_TLS_quality-factor}
to the contribution from direct ionization, Eq.~\eqref{eq:Q-estimation_clean-limit_final-estimate}:
\begin{equation}
\frac{Q^{\text{clean}}_{\text{ioniz}}}{Q^{\text{clean}}_{\text{TLS}}}\sim b^{2}\left(\frac{\varepsilon_{C}}{\Delta_{0}}\right)^{2}\frac{\omega}{T_{c}}\ll1.\label{eq:clean-case_ionization-to-TLS_comparison}
\end{equation}
Although the quantitative accuracy of this estimate is limited due
to the approximate nature of both individual expressions, the ratio
is expected to be well below unity---even in the worst-case scenario
with $b\sim10$, $\varepsilon_{C}/\Delta_{0}\sim0.1$, and $\omega\sim0.1\,T_{c}$.
This supports the general picture presented in Supplementary Materials
Fig.~\ref{fig:nonequil-qp-dissip_types_qualitative-diag}: direct
ionization dominates the loss budget unless the frequency is sufficiently
low. Moreover, Eq.~\eqref{eq:clean-case_ionization-to-TLS_comparison}
also implies that under most conditions, quasiparticle TLS dissipation
is even more likely to be overshadowed by the conventional channels,
such as dielectric loss.

The physics of TLS-like subsystems in superconducting devices---and,
notably, the dependence of dissipation on applied power---has received
considerable attention in the literature~\citep{Muller19}. In particular,
the growth and eventual saturation of the quality factor with increasing
power is considered evidence of TLS-type dissipation (see Refs.~\citep{Wang2009,hung2024}
as examples). Indeed, the response of an individual TLS, Eq.~\eqref{eq:single-dipole_absorbed-power},
generalizes to the weakly nonlinear regime by substituting $(\omega-\omega_{0})^{2}\mapsto(\omega-\omega_{0})^{2}+\left(\boldsymbol{p},\boldsymbol{E}\right)^{2}\Gamma_{2}/\Gamma_{1}$
in the denominator (see, e.g., Ref.~\citep[Eq. (5)]{faoro2012}),
resulting in growth and saturation of dissipation with increasing
power. For completeness, we note that quasiparticle ionization can
also exhibit a similar power dependence due to the nontrivial steady-state
distribution created by the pump; see the following subsec.~\ref{subsec:Dissipation_deloc-qps}.

However, the power absorbed by an \emph{ensemble} of TLSs strongly
depends on their spectral density $n(\omega)$, even in the linear
regime described above. The profile of $n(\omega)$ is the key difference
between the quasiparticle-induced TLSs discussed here and the standard
tunneling model~\citep[Eq. (22)]{Muller19}. The latter assumes a
weak frequency dependence of the spectral density, $n(\omega)\sim\omega^{\mu}$
with $\mu=0-0.3$, as dictated by the atomic nature of the TLSs~\citep[Eqs. (9) and (10)]{Faoro15}.
In contrast, quasiparticle TLSs exhibit a strongly $\omega$-dependent
spectral density {[}see Eq.~\eqref{eq:clean-case_TLS_spectral-density}{]}.
Therefore, the standard tunneling model~\citep[Eq. (22)]{Muller19}
cannot be directly applied to analyze the power dependence of the
quality factor. Instead, further theoretical and experimental effort
is required to characterize the nonlinear dissipation due to quasiparticle
TLSs.

\subsubsection{Dissipation due to delocalized quasiparticles\label{subsec:Dissipation_deloc-qps}}

Having discussed the main possibilities involving localized quasiparticles,
we now turn to a more straightforward scenario: either due to a high
effective quasiparticle temperature $T_{\text{eff}}$ or a sufficiently
strong external source, quasiparticles may occupy delocalized states
in the bulk of the excitation spectrum in the steady state (see subsec.~\ref{subsec:QP-distrib-func-and-concentation}).

In the dirty limit, such quasiparticles contribute to the dissipative
conductivity via an expression analogous to Eq.~\eqref{eq:sigma-1_MT-like-answer_diry-case},
with $\nu_{\text{loc}}(E)$ replaced by $\nu_{\text{AG}}(E)$, the
appropriate density of initial states. Neglecting the frequency dependence
of the final density of states immediately leads to the Mattis-Bardeen-type
result~\eqref{eq:sigma-1_Mattis-Bardeen}. A more careful evaluation
in the case of a quasi-equilibrium distribution function, Eq.~\eqref{eq:deloc-qps_quasi-equilib_distrib-function},
yields a numerical prefactor for Eq.~\eqref{eq:sigma-1_Mattis-Bardeen}
that depends on $\eta$, $T_{\mathrm{eff}}/\Delta_{0}$, and $\omega/\Delta_{0}$~\citep[see Eqs. (58)--(63)]{Fominov2011}. 
The estimate~\eqref{eq:sigma-1_Mattis-Bardeen} is, in fact, also
applicable in the clean limit~\citep{Fominov2011,Mattis58}.

For the dirty limit and the case when $\omega$ is large enough that
the final state lies beyond the BCS coherence peak (regardless of
$T_{\mathrm{eff}}$), the quality factor is estimated using Eqs. (59)
and (62) of Ref.~\citep{Fominov2011} for $\sigma_{1}$: 
\begin{equation}
	Q^{\text{dirty}}_{\text{MT}}\approx\frac{1}{\alpha}\frac{1}{\sqrt{2}}\left(\frac{\omega}{\Delta}\right)^{3/2}\frac{1}{n_{\text{qp}}\xi^{3}}\,\frac{\hbar\sigma_{2}\xi}{e^{2}},\label{eq:quality-factor_Mattis-Bardeen-deloc-qps_dirty}
\end{equation}
where we have used $\sigma_{0}=2\nu_{0}e^{2}D$ and assumed the 3D
limit $d\gg\xi$ for simplicity.

Crucially, the value of $n_{\text{qp}}\xi^{3}$ for delocalized
quasiparticles is difficult to predict theoretically, as the recombination
rate balancing the external quasiparticle generation exhibits a much
smoother energy dependence and is highly sensitive to the specific
microscopic details.

One can, however, use Eq.~\eqref{eq:quality-factor_Mattis-Bardeen-deloc-qps_dirty}
to estimate the concentration of delocalized quasiparticles required
to account for the observed dissipation. For the best Al cavities
in Fig.~\ref{fig1}, this yields $n_{\text{qp}}\xi^{3}=4.3$.
This value is roughly two orders of magnitude larger than the estimate
in Eq.~\eqref{eq:stationary-concentration_main-eq_2} for localized
quasiparticles, echoing the result obtained for localized quasiparticles
in the clean limit, Eq.~\eqref{eq:Q-estimation_clean-limit_final-estimate},
despite the limited accuracy of that estimate. One thus concludes
that while modern devices have made significant progress in suppressing
excess quasiparticles, their localization is not yet achieved, even
for the best-performing cavities represented on the right side of
Fig.~\ref{fig1} of the main text. This is inline with the conclusions
made in Refs.~\citep{Fischer2025,gurra2025}, see also sec.~\ref{subsec:Loc-deloc_recombination_limit}.

We also note that the evolution of the distribution function of delocalized
quasiparticles with applied power has recently been described comprehensively
from a theoretical standpoint~\citep{Catelani2019,Fischer23}, paving
the way for a more accurate understanding of the observed power dependence
of the quality factor. In particular, Ref.~\citep{Catelani2019}
demonstrates that the specific form of this dependence is sensitive
to both the properties of the quasiparticle-generating agent and material
parameters, making a simple universal description unlikely.

\subsubsection{Other possible loss mechanisms due to nonequilibrium quasiparticles at low frequencies\label{subsec:Localize-qps_low-freq_possible-loss-channels}}

The theoretical approach of subsecs.~\ref{subsec:Localized-states_direct-ionization_dirty-limit}~and~\ref{subsec:Localized-states_direct-ionization_clean-limit}
considers only nonmagnetic impurities, since the best-performing devices
are typically characterized by negligible magnetic disorder, both
on the surface and in the bulk~\citep[Intr.]{DeGraaf20}. However,
a finite concentration~$n_{s}$ of \emph{magnetic} impurities has
two main effects:

\emph{i)~}it contributes to the spatial variation of the order parameter,
thus influencing the depairing parameter~$\eta$ and the parameters
$\nu_{T},\varepsilon_{T}$ of the localized DoS tail (clean limit:
Ref.~\citep{DeGraaf20}, dirty limit: Ref.~\citep{Skvortsov13}).
Our analysis is not sensitive to the exact values of these parameters,
hence no qualitative change is expected.

\emph{ii)~}A localized magnetic impurity hosts a Yu-Shiba-Rusinov
state below the superconducting gap~(see Refs.~\citep{Fominov2011,babkin2022a}
and references therein). At sufficiently large~$n_{s}$, these states
merge into an umpurity band~\citep{Fominov2011,fominov2016}, whose
edges are further smeared by optimal disorder fluctuations hosting
localized states~\citep{fominov2016}. The YSR states can be filled
by quasiparticles, complicating the analysis of the resulting dissipation.
In particular, Ref.~\citep{Fominov2011} explores the dissipation
due to transitions within the impurity band, as well as between the
impurity band and the main quasiparticle continuum. This work identifies
the frequency range within which neither are possible, implying a
substantial suppression of dissipation. However, in light of the complicated
structure of the DoS~\citep{fominov2016}, multiple other regimes
of suppressed dissipation can be imagined.

As a result, a comprehensive study of the role and the relevance of
magnetic impurities in the dissipation of modern superconducting devices
remains an open question.

While most modern devices arguably operate in the regime $\omega>\varepsilon_{T}$,
where dissipation is of the Mattis-Bardeen type {[}Eq.~\eqref{eq:sigma-1_Mattis-Bardeen}{]},
certain cases feature frequencies low enough to render alternative
mechanisms relevant~\citep{DeGraaf20}, with quasiparticle TLSs discussed
above being one example. It is also possible that, in the dirty limit,
the number of bound states in a typical local dip of the order parameter
exceeds one~\citep{Meyer_private}. However, as pointed out in Ref.~\citep{DeGraaf20},
direct transitions between these bound states are expected to be inefficient
in the dirty case due to their small transition dipole moments.

This nevertheless leaves room for several other dissipation mechanisms,
illustrated in Supplementary Materials Figs.~\ref{fig:qp_dissip-mechanisms}\textbf{e} and~\ref{fig:nonequil-qp-dissip_types_qualitative-diag}.
If the actual level spacing between bound states within a well is small enough,
the local spectrum may effectively behave as continuous, resulting
in Mattis-Bardeen-like dissipation \emph{within} each individual well.
Alternatively, slow relaxation rates of localized quasiparticles (see
subsec.~\ref{subsec:QP-distrib-func-and-concentation}) may favor
nonequilibrium dynamics within a single well, resembling the mechanism
behind the enhanced dissipative response in thin superconducting films
in the mixed state~\citep{PAF2023}. Due to the altered quasiparticle
distribution function in such a scenario, the power dependence of
the quality factor is expected to become even more pronounced.

Finally, another intriguing possibility is photon-assisted variable-range
hopping between different local dips of the order parameter (see Supplementary
Materials Fig.~\ref{fig:qp_dissip-mechanisms}\textbf{f}), a mechanism
that, to our knowledge, has not been systematically studied. While
the problem bears some resemblance to its semiconductor analog~\citep{Mott1970},
the strong energy dependence of the density of localized states, Eqs.~\eqref{eq:localization-properties_clean-limit}--\eqref{eq:localization-properties_dirty-limit},
must be taken into account.

\subsection{Dissipation by an isolated Vortex}\label{sec:dissipation_by_vortex}

The inverse quality factor \( 1/Q \) of a resonator in the presence of a low magnetic field is determined by contributions from individual vortices that penetrate the superconducting film and become pinned by disorder. In this low-field regime, variations in the effective superfluid stiffness \( \Theta(B) \) are negligible. Thus, the dissipation contribution to $\frac{1}{Q} = \frac{\sigma_1}{\sigma_2}$
is governed by the behavior of the real part of the conductivity \( \sigma_1(\omega) \). The complex sheet conductivity of a superconductor in the mixed state with pinned vortices is given by (see e.g., Ref.~\cite{PAF2023,PAFcorr}): 
\begin{equation}
\sigma^{\Box}(\omega) = \frac{i d}{\omega \mu_0} \frac{1}{\lambda_L^2 + \frac{\Phi_0 B d}{ \mu_0 (k_0 - i\omega\eta)}}.
\label{sigma1}
\end{equation}
In the above expression, \( \lambda_L \) is the London penetration depth,  \( d \) is the film thickness,  \( k_0 \) is the curvature of the vortex pinning potential,  and \( \eta_v \) is the vortex friction coefficient, determining the friction force \( f_v = \eta_v v \) acting on a vortex moving at velocity \( v \). 
The friction coefficient \( \eta_v \) is related to the bulk DC conductivity in the flux-flow regime (at currents well above the depinning threshold) via \( \sigma_{ff} = \frac{ \eta_v}{B \Phi_0 d} \). It is further connected to the normal-state conductivity \( \sigma_n \) by  
\[
\eta_v = \zeta \Phi_0 H_{c2} \mu_0 \sigma_n d,
\]
where \( \zeta \sim 1 \) is a numerical factor.

Expanding the real part of \( \sigma \) from Eq.~(\ref{sigma1}) in the small \( B \) limit, we find:
\begin{equation}
\sigma_1^{\Box} = \frac{\Phi_0 B \eta_v d^2}{\mu_0^2 \lambda_L^4 k_0^2 \zeta} 
\sim \frac{\sigma_n B \, \Theta}{\Phi_0 j_c^2 \lambda_L^2},
\label{sigma2}
\end{equation}
where \( j_c \) is the critical current density of the material, \( \Theta = \Phi_0^2 d / (4\pi^2 \mu_0 \lambda_L^2) \) is the superfluid stiffness, and we have used the approximate relations \( \Phi_0 j_c d  \approx k_0 \xi_0 \) and \( H_{c2} \approx \Phi_0 / 2\pi \mu_0 \xi_0^2 \). 
Using Eq.~(\ref{sigma2}) together with the standard expression for the pair-breaking current density \( j_0 = \Phi_0 / (3 \sqrt{3} \pi \mu_0 \lambda_L^2 \xi_0 ) \), we can express the vortex-induced contribution to the inverse quality factor as:
\begin{equation}
    \frac{1}{Q_{\text{Vortex}}} \sim 10 \, \frac{\sigma_n d}{\sigma_2^\Box} \frac{B}{\mu_0 H_{c2}} \left( \frac{j_0}{j_c} \right)^2.
    \label{Q-v}
\end{equation}
Considering a residual field strength $B_0 \sim 10^{-7}$ T, the factor $B_0/ \mu_0 H_{c2} \sim 10^{-6} - 10^{-8}$ for different materials. The last factor in Eq.~(\ref{Q-v}) is very much material-dependent and can reach $10^4-10^6$. 
This non-universality of the prefactor of $\sigma_2$ in Eq.~(\ref{Q-v}) is also reflected in the fact that highly disordered films exhibit strong vortex pinning, which limits vortex-induced dissipation, in contrast to low-disorder films, where weaker pinning results in a higher \( j_0/j_c \) ratio.
According to Eq.(\ref{Q-v}) one could expect rather low $Q$-factor due to vortices in low-disorder films; however, i) at low $T$ vortices are energetically unfavorable at very low external fields $B \ll \Phi_0/w^2$ where $w \sim $ few microns is typical width of a resonator stripe \cite{Stan04}, and ii) precisely due to weak pinning in low-disordered material, vortices are not pinned in such a narrow stripes. Thus they are expelled out of a stripe to decrease energy.


\begin{figure*}
\begin{centering}
\includegraphics[scale=0.69]{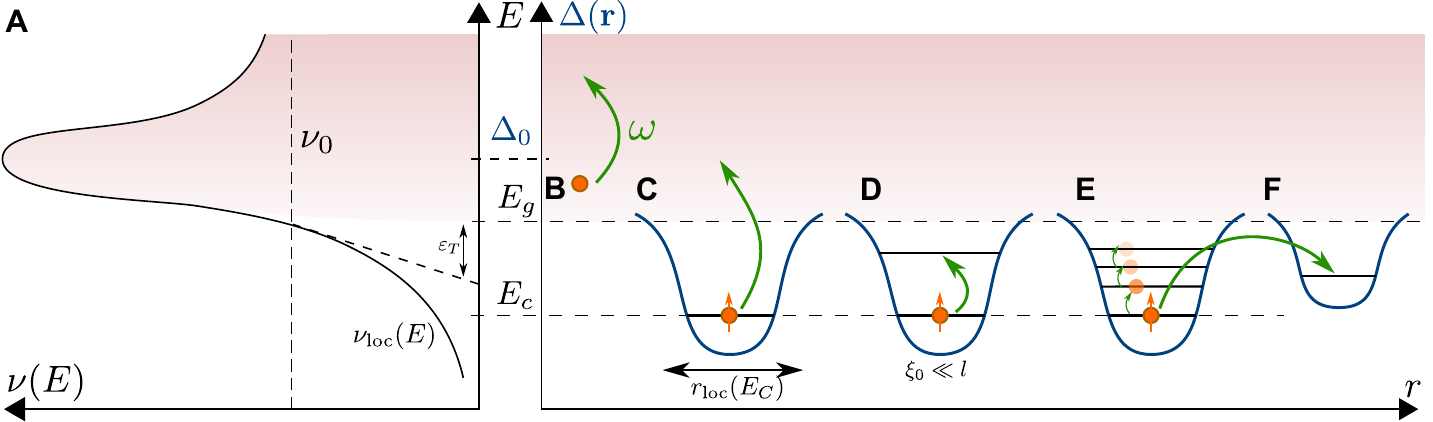}
\par\end{centering}
\caption{\textbf{Dissipation mechanisms induced by nonequilibrium quasiparticles.}
\textbf{a},~Sketch of the quasiparticle density of states $\nu(E)$
as a function of energy near the superconducting gap edge. $\nu_{0}$~denotes
the normal-state density of states per spin projection at the Fermi
level; $\Delta_{0}$~is the superconducting gap in the clean limit,
and $E_{g}\approx\Delta_{0}(1-\eta^{2/3})^{3/2}$ marks the mobility
edge separating the delocalized states (pink) from the localized tail.
The difference between $E_{g}$ and $\Delta_{0}$ arises from disorder-induced
smearing of the BCS coherence peak, characterized by the depairing
parameter $\eta$~\citep{Abrikosov1960,Larkin72}. $\nu_{\text{loc}}(E)$
represents the subgap tail of localized quasiparticle states, with
a characteristic energy scale $\varepsilon_{T}\ll\Delta_{0}$, and
$E_{C}$ denotes the energy threshold below which quasiparticle recombination
becomes inefficient~\citep{Bespalov16}.
\textbf{ b--f},~Schematic
representation of possible dissipation mechanisms involving nonequilibrium
quasiparticles, approximately ordered by decreasing dissipation strength
(see also Fig.~\ref{fig:nonequil-qp-dissip_types_qualitative-diag}):
\textbf{b},~Standard Mattis-Bardeen-like transitions in the continuum
of delocalized states\citep{Mattis58,Fominov2011}, subsec.~\ref{subsec:Dissipation_deloc-qps};
\textbf{c},~Ionization of a bound quasiparticle state in a local
well of suppressed superconducting order parameter (blue) with characteristic
size $r_{\text{loc}}(E_{C})$, subsecs.~\ref{subsec:Localized-states_direct-ionization_dirty-limit}
and~\ref{subsec:Localized-states_direct-ionization_clean-limit};
\textbf{d},~Coherent Rabi oscillations between two bound states within
a single well in the clean limit, subsec.~\ref{subsec:Localized-qp_clean-limit_coherent-TLS};
\textbf{e,f},~Semiclassical dynamics involving multiphoton absorption
and photon-assisted variable-range hopping between different localized
wells, potentially in nonequilibrium regimes, subsec.~\ref{subsec:Localize-qps_low-freq_possible-loss-channels}.\label{fig:qp_dissip-mechanisms}}
\end{figure*}

\begin{figure*}
\begin{centering}
\includegraphics[scale=0.7]{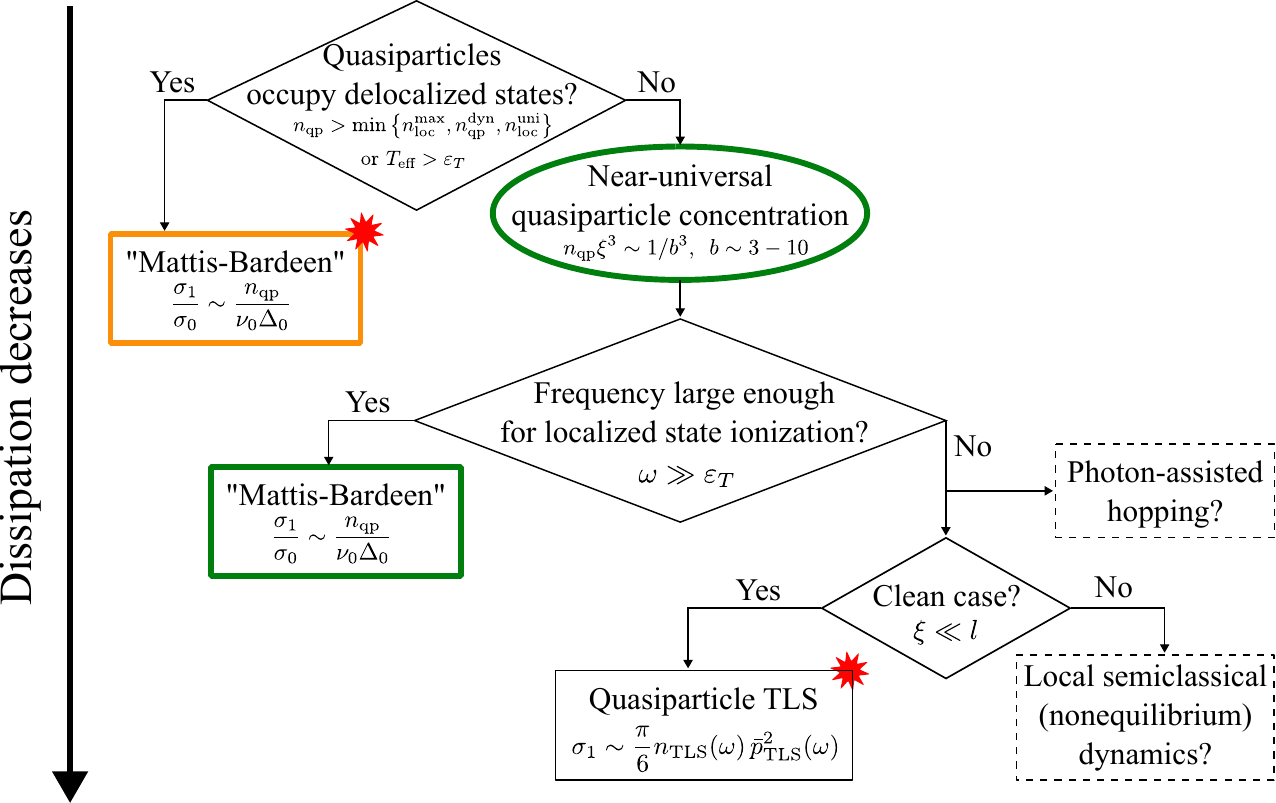}
\par\end{centering}
\caption{\textbf{Classification of the main dissipation mechanisms due to nonequilibrium
quasiparticles.} Green shading indicates the dominant dissipation
mechanism in dirty superconducting thin films (left side of Fig.~\ref{fig1}, main text), as discussed in subsec.~\ref{subsec:Localized-states_direct-ionization_dirty-limit}.
Orange indicates the dominant mechanism in clean films and 3D cavities
(right side of Fig.~\ref{fig1}, main text; see subsec.~\ref{subsec:Dissipation_deloc-qps}).
A red star marks those mechanisms for which a power-dependence analysis
of the quality factor is available. Dashed gray boxes denote conjectured
and / or poorly understood dissipation mechanisms. \label{fig:nonequil-qp-dissip_types_qualitative-diag}}
\end{figure*}

\begin{figure}[ht!]
\centering
 \includegraphics[width=1\linewidth]{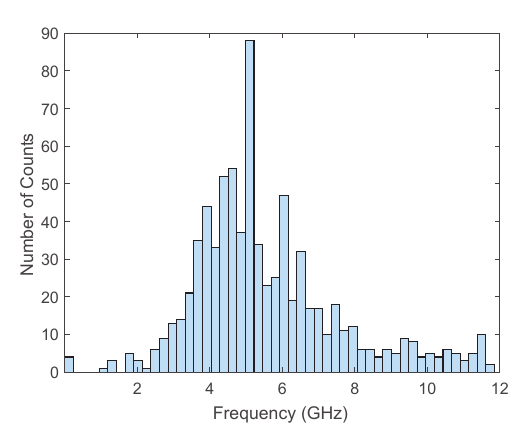}
\caption{\textbf{Frequency histogram of the datasets.} Histogram of the frequencies corresponding to the datasets shown in Fig.~\ref{fig1}.} \label{fig:extended_fig1}
\end{figure}

\clearpage

\linespread{1.0}

{\centering
\begin{longtable*}[H]{ccccccccccc} 
\caption{\textbf{Table of resonator and qubit parameters for the data shown in Fig.~\ref{fig1} and Supplementary Materials Fig.~\ref{fig:extended_fig1}.} 
The resonators have either planar geometry (thickness $d$) or are 3D cavities. Quality factors are measured at low temperatures (typically $T \sim 20$ mK) and in the low-photon-number regime, at frequency $f$. Material properties such as critical temperature $T_c$ and sheet normal-state resistance $R_{\square}$ (measured above the superconducting transition) depend on the material, its purity (residual-resistivity ratio, RRR), and the film thickness. Resistance values reported only at room temperature are labeled (RT). 
In the dirty limit, the sheet kinetic inductance can be calculated using $L_K = \hbar R_{\square} / (\pi \Delta)$, where $\Delta = 1.76\,k T_c$ for conventional BCS superconductors. $L_K$ can also be inferred from electromagnetic simulations, from fits to the Mattis-Bardeen theory~\cite{Mattis58} based on the temperature dependence of the resonance frequency $f(T)$, or from fits to the plasmon dispersion relation~\cite{Charpentier2024}. 
In the clean limit, we use tabulated values for the London penetration depth $\lambda$ for bulk superconductors, with thickness corrections applied when available. Data that are partially missing but can be reasonably estimated from references are marked with a star (*). The estimation of $\lambda$ and $L_K$ for cleaner superconductors (e.g., Nb, Al, Ta) is often complicated by the absence of transport characterization and is therefore subject to larger uncertainties. The $Q_i$ value of the grAl transmon from Ref.~\cite{Winkel_grAl_Transmon_2020} is obtained as the mean of the different measurement runs for the same device reported in Table II of that work.}\label{Table1}\\
\toprule
Material & Ref & $d$ (nm) & $R_{\square}~(\mathrm{\Omega/\square})$ & $T_c$ (K) & $f$ (GHz) & $\lambda$ (nm) & $L_K~(\mathrm{pH / \square})$ & $\sigma_2~(\times 10^8~\mathrm{S/m})$ & $\alpha$ & $Q_i~(\times 10^{5})$\\ 
\midrule
\endfirsthead
\multicolumn{10}{c}%
{\tablename\ \thetable\ -- \textit{Continued from previous page}} \\
\toprule
Material & Ref & $d$ (nm) & $R_{\square}~(\mathrm{\Omega/\square})$ & $T_c$ (K) & $f$ (GHz) & $\lambda$ (nm) & $L_K~(\mathrm{pH / \square})$ & $\sigma_2~(\times 10^8~\mathrm{S/m})$  & $\alpha$ & $Q_i~(\times 10^{5})$\\ 
\midrule
\endhead
\bottomrule \multicolumn{10}{r}{\textit{Continued on next page}} \\
\endfoot
\bottomrule
\endlastfoot

Al & \cite{deVisser2014} & 60 & 0.15 & 1.17 & 5.289 & 92 &  & 28 & & 25 \\
Al & \cite{Burnett2018} & 150 &  &  & 5.244 & 65 \cite{Lopez2023}&  & 57 & & 7.9 \\
Al & \cite{Megrant2012} & 100 &  &  & 3.83 & 80 \cite{Lopez2023} &  & 51.6 & $8.2 \, 10^{-2}$ & 1.6 \\
Al & \cite{Megrant2012} & 100 &  &  & 6.129 & 80 \cite{Lopez2023} &  & 32 & $1.78 \, 10^{-2}$ & 3 \\
Al & \cite{Megrant2012} & 100 &  &  & 3.81 & 80 \cite{Lopez2023} &  & 51.9 & $8.2 \, 10^{-2}$ & 6.6 \\
Al & \cite{Megrant2012} & 100 &  &  & 6.08 & 80 \cite{Lopez2023} &  & 32.5 & $1.78 \, 10^{-2}$ & 7.2 \\
Al & \cite{Megrant2012} & 100 &  &  & 4.97 & 80 \cite{Lopez2023} &  & 39.8 & $4.3 \, 10^{-2}$ & 5.3 \\
Al & \cite{Megrant2012} & 100 &  &  & 6.12 & 80 \cite{Lopez2023} &  & 32.3 & $1.78 \, 10^{-2}$ & 7.6 \\
Al & \cite{Megrant2012} & 100 &  &  & 3.77 & 80 \cite{Lopez2023} &  & 52.5 & $8.2 \, 10^{-2}$ & 7.5 \\
Al & \cite{Megrant2012} & 100 &  &  & 6.125 & 80 \cite{Lopez2023} &  & 32.3 & $1.78 \, 10^{-2}$ & 8 \\
Al & \cite{Megrant2012} & 100 &  &  & 3.87 & 80 \cite{Lopez2023} &  & 51.1 & $8.2 \, 10^{-2}$ & 11.5 \\
Al & \cite{Megrant2012} & 100 &  &  & 6.12 & 80 \cite{Lopez2023} &  & 32.3 & $1.78 \, 10^{-2}$ & 9.2 \\
Al & \cite{Megrant2012} & 100 &  &  & 3.76 & 80 \cite{Lopez2023} &  & 52.6 & $8.2 \, 10^{-2}$ & 11 \\
Al & \cite{Megrant2012} & 100 &  &  & 6.12 & 80 \cite{Lopez2023} &  & 32.3 & $1.78 \, 10^{-2}$ & 17.2 \\
Al & \cite{Earnest2018} & 100 &  &  & 4.487 & 80 \cite{Lopez2023} &  & 44 & $1.8 \, 10^{-2}$ & 3.47 \\
Al & \cite{Earnest2018} & 100 &  &  & 4.501 & 80 \cite{Lopez2023} &  & 44 & $1.8 \, 10^{-2}$ & 7.21 \\
Al & \cite{Earnest2018} & 100 &  &  & 4.512 & 80 \cite{Lopez2023} &  & 44 & $1.8 \, 10^{-2}$ & 7.17 \\
Al & \cite{Earnest2018} & 100 &  &  & 4.506 & 80 \cite{Lopez2023} &  & 44 & $1.8 \, 10^{-2}$ & 9.85 \\
Al & \cite{Richardson2016} & 100 &  &  & 5.606 & 80 \cite{Lopez2023} &  & 35 & & 7.5 \\ 
Al & \cite{Wang2009} & 120 &  &  & 6.7 & 80 \cite{Lopez2023} &  & 29.5 & $5.5 \, 10^{-2}$ & 0.6\\ 
Al & \cite{Wang2009} & 120 &  &  & 7 & 80 \cite{Lopez2023} &  & 28.2 & $5.5 \, 10^{-2}$ & 0.6\\ 
Al & \cite{Wang2009} & 120 &  &  & 7.1 & 80 \cite{Lopez2023} &  & 27.8 & $2.8 \, 10^{-2}$ & 11\\ 
Al & \cite{Wang2009} & 120 &  &  & 7 & 80 \cite{Lopez2023} &  & 28.2 & $5.5 \, 10^{-2}$ & 0.7\\ 
Al & \cite{Wang2009} & 120 &  &  & 7 & 80 \cite{Lopez2023} &  & 28.2 & $3.5 \, 10^{-2}$ & 11\\ 
Al & \cite{Wang2009} & 120 &  &  & 7.1 & 80 \cite{Lopez2023} &  & 27.8 & $1.8 \, 10^{-2}$ & 17\\ 
Al & \cite{Biznarova2024}~ & 500 &  &  & Fig.2 & 52 \cite{Lopez2023} &  & & $ 8.1 \, 10^{-3}$ & Fig.2\\ 
Al & \cite{Biznarova2024} & 300 &  &  & Fig.2 & 52 \cite{Lopez2023} &  & & $ 8.1 \, 10^{-3}$ & Fig.2\\ 
Al & \cite{Biznarova2024} & 150 &  &  & Fig.2 & 65 \cite{Lopez2023} &  & & $ 1.0 \, 10^{-2}$ & Fig.2\\
Al & \cite{Mahuli2025} & 100 &  &  & 5.3 & 80 &  & 37.3  &  & 34\\
Al Transmon & \cite{Biznarova2024} & 500 &  &  & Tab. II & 52 \cite{Lopez2023} &  & &  & Tab. II\\ 
Al Transmon & \cite{Biznarova2024} & 300 &  &  & Tab. II & 52 \cite{Lopez2023} &  & &  & Tab. II\\ 
Al Transmon & \cite{Biznarova2024} & 150 &  &  & Tab. II & 65 \cite{Lopez2023} &  & &  & Tab. II\\
Al Transmon & \cite{Wang_Transmon_2022} & 120 &  &  & Tab. I & 63 \cite{Lopez2023}&  &  &  & Tab. I \\ 
Al Transmon & \cite{Kurter_NbN_Transmon2022} & 200 &  &  & 5* & $\sim$ 100* &  &  &  & Fig. 2b\\
Al Transmon & \cite{Mahuli2025} & 100 &  &  & 5.7 & 80 &  & 34.7  &  & 36.9\\
Al susp. strips & \cite{Krayzman2024} & 2000 &  &  & 5* & 52 & & 93.7* & $ 3 \, 10^{-5}$ & 440 \\ 
Al susp. strips & \cite{Krayzman2024} & 2000 &  &  & 5* & 52 & & 93.7* & $ 3 \, 10^{-5}$ & 480 \\ 
Al susp. strips & \cite{Krayzman2024} & 600 &  &  & 5* & 52 & & 93.7* & $ 3 \, 10^{-5}$ & 370 \\ 
Al susp. strips & \cite{Krayzman2024} & 800 &  &  & 5* & 52 & & 93.7* & $ 3 \, 10^{-5}$ & 370 \\ 
Al susp. strips & \cite{Krayzman2024} & 600 &  &  & 5* & 52 & & 93.7* & $ 3 \, 10^{-5}$ & 1080 \\ 
Al susp. strips & \cite{Krayzman2024} & 600 &  &  & 5* & 52 & & 93.7* & $ 3 \, 10^{-5}$ & 1270 \\ 
Al susp. strips & \cite{Krayzman2024} & 600 &  &  & 5* & 52 & & 93.7* & $ 3 \, 10^{-5}$ & 96 \\ 
Al susp. strips & \cite{Krayzman2024} & 800 &  &  & 5* & 52 & & 93.7* & $ 3 \, 10^{-5}$ & 370 \\ 
\midrule
Al (3D) & \cite{Reagor2013} & bulk &  & 1.18 & 9.613 & 65 & & 31.5 &   & 51 \\
Al (3D) & \cite{Reagor2013} & bulk &  & 1.18 & 9.464 & 65 & & 31.6 &   & 56 \\
Al (3D) & \cite{Reagor2013} & bulk &  & 1.18 & 9.478 & 65 & & 31.6 &  & 48 \\
Al (3D) & \cite{Reagor2013} & bulk &  & 1.18 & 7.69 & 65 & & 39 &   & 24 \\
Al (3D) & \cite{Reagor2013} & bulk &  & 1.18 & 11.448 & 65 & & 26 &  & 10 \\
Al (3D) & \cite{Reagor2013} & bulk &  & 1.18 & 11.416 & 65 & & 26 &  & 1.5 $10^3$ \\
Al (3D) & \cite{Reagor2013} & bulk &  & 1.18 & 11.45 & 65 & & 26 &  & 150 \\
Al (3D) & \cite{Reagor2013} & bulk &  & 1.18 & 9.455 & 52 & & 49.5 & & 420 \\
Al (3D) & \cite{Reagor2013} & bulk &  & 1.18 & 9.481 & 52 & & 49.4 & & 430 \\
Al (3D) & \cite{Reagor2013} & bulk &  & 1.18 & 9.481 & 52 & & 49.4 & & 690 \\
Al (3D) & \cite{Reagor2013} & bulk &  & 1.18 & 7.7 & 52 & & 60.8 & & 310 \\
Al (3D) & \cite{Reagor2013} & bulk &  & 1.18 & 11.448 & 52 & & 40.9 & & 140 \\
Al (3D) & \cite{Reagor2013} & bulk &  & 1.18 & 11.417 & 52 & & 41.0 & & 6.09 $10^3$ \\
Al (3D) & \cite{Reagor2013} & bulk &  & 1.18 & 11.419 & 52 & & 41.0 & & 320 \\
Al (3D) & \cite{Reagor2013} & bulk &  & 1.18 & 11.44 & 52 & & 40.9 & & 7.4 $10^3$ \\
Al (3D) & \cite{Reagor2013} & bulk &  & 1.18 & 11.442 & 52 & & 40.9 & & 5.2 $10^3$ \\
Al alloy (3D) & \cite{Reagor2013} & bulk &  & 1.18 & 9.513 & 194 & & 3.53 & & 51 \\ 
Al alloy (3D) & \cite{Reagor2013} & bulk &  & 1.18 & 9.450 & 194 & & 3.56 & & 32 \\
Al (3D) & \cite{Kudra2020} & bulk &  & 1.18 & 7.43 & 62 & & 44.3 & 5 $10^{-5}$ & 830 \\
Al (3D) & \cite{Kudra2020} & bulk &  & 1.18 & 7.417 & 62 & & 44.4 & 5 $10^{-5}$ & 660 \\
Al (3D) & \cite{Kudra2020} & bulk &  & 1.18 & 7.425 & 62 & & 44.3 & 5 $10^{-5}$ & 810 \\
Al (3D) & \cite{Kudra2020} & bulk &  & 1.18 & 7.417 & 62 & & 44.4 & 5 $10^{-5}$ & 930 \\
Al (3D) & \cite{Kudra2020} & bulk &  & 1.18 & 7.427 & 62 & & 44.3 & 5 $10^{-5}$ & 790 \\
Al (3D) & \cite{Kudra2020} & bulk &  & 1.18 & 7.428 & 62 & & 44.3 & 5 $10^{-5}$ & 910 \\
Al (3D) & \cite{Kudra2020} & bulk &  & 1.18 & 7.427 & 62 & & 44.3 & 5 $10^{-5}$ & 820 \\
Al (3D) & \cite{Kudra2020} & bulk &  & 1.18 & 6.476 & 62 & & 50.8 & 5 $10^{-5}$ & 860 \\
Al (3D) & \cite{Kudra2020} & bulk &  & 1.18 & 5.478 & 62 & & 60.1 & 5 $10^{-5}$ & 940 \\
Al (3D) & \cite{Kudra2020} & bulk &  & 1.18 & 4.5 & 62 & & 73.2 & 5 $10^{-5}$ & 1.15 $10^3$ \\
Al (3D) & \cite{Kudra2020} & bulk &  & 1.18 & 5.932 & 62 & & 55.5 & 5 $10^{-5}$ & 300 \\
Al (3D) & \cite{Kudra2020} & bulk &  & 1.18 & 7.437 & 62 & & 44.3 & 5 $10^{-5}$ & 1.01 $10^3$ \\
Al (3D) & \cite{Kudra2020} & bulk &  & 1.18 & 5.928 & 62 & & 55.5 & 5 $10^{-5}$ & 750 \\
Al (3D) & \cite{Kudra2020} & bulk &  & 1.18 & 5.923 & 62 & & 55.6 & 5 $10^{-5}$ & 840 \\
Al alloy (3D) & \cite{Kudra2020} & bulk &  & 1.18 & 5.976 & 189 & & 5.933 &  & 50 \\
Al alloy (3D) & \cite{Kudra2020} & bulk &  & 1.18 & 5.929 & 189 & & 5.98 &  & 120 \\
Al alloy (3D) & \cite{Kudra2020} & bulk &  & 1.18 & 5.939 & 189 & & 5.97 &  & 70 \\
Al (3D) & \cite{Chakram21} & bulk &  &  & 6.389 & 235 & & 3.58 & $4.61 \, 10^{-5}$ & 250 \\
Al (3D) & \cite{Chakram21} & bulk &  &  & 6.75 & 37.6 & & 132 & $1.89 \, 10^{-5}$ & 100 \\ 
Al alloy (3D) & \cite{Esmenda2025} & bulk &  & 1.18 & 2.5123 & 194 & & 13.39 &  & 260 \\
Al alloy (3D) & \cite{Esmenda2025} & bulk &  & 1.18 & 2.8156 & 194 & & 11.95 &  & 220 \\
Al alloy (3D) & \cite{Esmenda2025} & bulk &  & 1.18 & 3.5571 & 194 & & 9.46 &  & 236 \\
Al alloy (3D) & \cite{Esmenda2025} & bulk &  & 1.18 & 4.0007 & 194 & & 8.41 &  & 247 \\
Al alloy (3D) & \cite{Esmenda2025} & bulk &  & 1.18 & 4.2245 & 194 & & 7.96 &  & 254 \\
Al alloy (3D) & \cite{Esmenda2025} & bulk &  & 1.18 & 4.2248 & 194 & & 7.96 &  & 88 \\
\midrule
grAl & \cite{Rotzinger2017} & 20 & 1153 & 1.6 & 4.04 & & 995 & 0.019 & 0.99 & 1.26 \\
grAl & \cite{Rotzinger2017} & 20 & 382 & 1.6 & 3.11 & & 329 & 0.077 & 0.98 & 0.37 \\
grAl & \cite{Zhang2019} & 20 &  & 1.7 & 4.05 & & 1200 & 0.019 & 1 & 0.2 \\
grAl & \cite{Zhang2019} & 20 &  & 1.7 & 4.05 & & 1200 & 0.021 & 1 & 0.13 \\
grAl & \cite{Zhang2019} & 20 &  & 1.75 & 3.88 & & 2000 & 0.012 & 1 & 0.09 \\
grAl & \cite{He2021} & 120 & 17 & 1.8 & 3.78 & & 20 & 0.17 & 0.56 & 1.0 \\
grAl & \cite{Rieger2023} & 20 &  &  & 5.0 & & 700~\cite{Gunzler2024} & 0.022 & & 0.773 \\
grAl & \cite{Rieger2023} & 20 &  &  & 5.644 & & 700~\cite{Gunzler2024} & 0.020 & 0.9 & 1.14 \\
grAl & \cite{Rieger2023} & 20 &  &  & 6.247 & & 700~\cite{Gunzler2024} & 0.018 & 0.9 & 1.34 \\
grAl & \cite{Rieger2023} & 20 &  &  & 6.848 & & 700~\cite{Gunzler2024} & 0.016 & 0.9 & 1.59 \\
grAl & \cite{Rieger2023} & 20 &  &  & 7.469 & & 700~\cite{Gunzler2024} & 0.015 & 0.9 & 1.70 \\
grAl & \cite{Rieger2023} & 20 &  &  & 8.273 & & 700~\cite{Gunzler2024} & 0.013 & 0.9 & 0.88 \\
grAl & \cite{Rieger2023} & 20 &  &  & 9.474 & & 700~\cite{Gunzler2024} & 0.012 & 0.9 & 1.25 \\
grAl & \cite{Rieger2023} & 20 &  &  & 10.538 & & 700~\cite{Gunzler2024} & 0.010 & 0.9 & 1.16 \\
grAl & \cite{Rieger2023} & 20 &  &  & 11.310 & & 700~\cite{Gunzler2024} & 0.010 & 0.9 & 1.86 \\ 
grAl & \cite{Roy2025} & 50 &  &  & \cite{Roy2025} Tab. I & & 150 &  &  & \cite{Roy2025} Tab. I \\ 
grAl & \cite{Khorramshahi2025} & 20 &  &  & 5.78 & & 670 & 0.020 &  & 1.19 \\ 
grAl & \cite{Khorramshahi2025} & 20 &  &  & 4.68 & & 670 & 0.025 &  & 2.58 \\ 
grAl & \cite{Khorramshahi2025} & 20 &  &  & 7.39 & & 530 & 0.020 &  & 3.25 \\ 
grAl & \cite{Khorramshahi2025} & 30 &  &  & 7.02 & & 260 & 0.029 &  & 1.29 \\
grAl & \cite{Khorramshahi2025} & 30 &  &  & 6.54 & & 260 & 0.031 &  & 1.77 \\
grAl & \cite{Khorramshahi2025} & 30 &  &  & 5.95 & & 180 & 0.049 &  & 2.76 \\   
grAl & \cite{Gupta2024} & 91 &  &  & 4.79 & & 50 & 0.073 &  &  26.7\\ 
grAl & \cite{Gupta2024} & 91 &  &  & 5.23 & & 50 & 0.069 &  &  23.0\\ 
grAl & \cite{Gupta2024} & 91 &  &  & 5.67 & & 50 & 0.061 &  &  18.3\\ 
grAl & \cite{Gupta2024} & 91 &  &  & 6.32 & & 50 & 0.055 &  &  23.8\\ 
grAl & \cite{Gupta2024} & 91 &  &  & 5.27 & & 70 & 0.047 &  &  16.7\\ 
grAl & \cite{Gupta2024} & 91 &  &  & 5.69 & & 70 & 0.043 &  &  15.9\\ 
grAl & \cite{Gupta2024} & 91 &  &  & 6.16 & & 70 & 0.040 &  &  16.1\\ 
grAl & \cite{Gupta2024} & 91 &  &  & 6.78 & & 70 & 0.037 &  &  13.3\\ 
grAl & \cite{Gupta2024} & 91 &  &  & 6.13 & & 120 & 0.023 &  &  18.9\\
grAl & \cite{Gupta2024} & 91 &  &  & 5.32 & & 120 & 0.027 &  &  14.5\\
grAl & \cite{Gupta2024} & 91 &  &  & 4.60 & & 320 & 0.012 &  &  5.2\\
grAl & \cite{Gupta2024} & 91 &  &  & 4.92 & & 320 & 0.011 &  &  5.6\\
grAl & \cite{Gupta2024} & 91 &  &  & 5.31 & & 320 & 0.010 &  &  5.8\\
grAl & \cite{Gupta2024} & 91 &  &  & 5.67 & & 320 & 0.009 &  &  4.6\\
grAl Transmon & \cite{Winkel_grAl_Transmon_2020} & 10 & 1800 (RT) & 1.9 & 7.57* & & 1100 &  &  & 3.96*  \\ 
grAl Transmon & \cite{Schon_grAl_Transmon_2020} & 20 & 7000* & 1.8 & 5.47 & & 270* &  &  & 1.16  \\
grAl Transmon & \cite{Schon_grAl_Transmon_2020} & 20 & 7000* & 1.8 & 8.50 & & 1891* &  &  & 2.13  \\
grAl Transmon & \cite{Schon_grAl_Transmon_2020} & 20 & 10000* & 1.8 & 7.93 & & 1350* &  &  & 1.69  \\
\midrule
Hf & \cite{Coiffard2020} & 125 & 7.3 & 0.385 & 5* & & 15.3 & 0.16 & & 3.25 \\
Hf & \cite{Coiffard2020} & 125 & 7.6 & 0.364 & 5* & & 17 & 0.15 & & 1.74 \\
Hf & \cite{Coiffard2020} & 125 & 8.25 & 0.403 & 5* & & 16.5 & 0.15 & & 4.05 \\
Hf & \cite{Coiffard2020} & 125 & 10.2 & 0.410 & 5* & & 20.6 & 0.12 & & 6.05 \\
Hf & \cite{Coiffard2020} & 125 & 8.35 & 0.435 & 5* & & 15.8 & 0.16 & & 5.15 \\
Hf & \cite{Coiffard2020} & 125 & 6.46 & 0.355 & 5* & & 13.2 & 0.19 & & 0.16 \\
Hf & \cite{Coiffard2020} & 125 & 8.11 & 0.395 & 5* & & 16.7 & 0.15 & & 1.9 \\ 
Hf & \cite{Li2025} & 250 &  & 0.25 & \cite{Li2025} Fig. 3 & & 4.8 &  & & \cite{Li2025} Fig. 3 \\
Hf & \cite{Li2025} & 250 &  & 0.25 & \cite{Li2025} Fig. 3 & & 5.3 &  & & \cite{Li2025} Fig. 3 \\ 
\midrule
a-InO & \cite{Charpentier2024}~TC002 & 40 & 1450 & 3.2 & 5.89 & & 590 & $11.4 \, 10^{-3}$ & 1 & 0.10 \\
a-InO & \cite{Charpentier2024}~TC014 & 40 & 1680 & 3.16 & 2.75 & & 700 & $10.6 \, 10^{-3}$ & 1 & 0.16 \\
a-InO & \cite{Charpentier2024}~TC014 & 40 & 1680 & 3.16 & 3.74 & & 700 & $15.2 \, 10^{-3}$ & 1 & 0.105 \\
a-InO & \cite{Charpentier2024}~TC003 & 40 & 2060 & 2.8 & 5.08 & & 910 & $8.6 \, 10^{-3}$ & 1 & 0.063 \\
a-InO & \cite{Charpentier2024}~TC003 & 40 & 2060 & 2.8 & 6.13 & & 910 & $7.1 \, 10^{-3}$ & 1 & 0.059 \\
a-InO & \cite{Charpentier2024}~TC003 & 40 & 2060 & 2.8 & 7.14 & & 910 & $6.1 \, 10^{-3}$ & 1 & 0.072 \\
a-InO & \cite{Charpentier2024}~TC040 & 40 & 2840 & 2.74 & 6.21 & & 1320 & $4.8 \, 10^{-3}$ & 1 & 0.10 \\
a-InO & \cite{Charpentier2024}~TC040 & 40 & 2840 & 2.74 & 7.82 & & 1320 & $3.8 \, 10^{-3}$ & 1 & 0.07 \\
a-InO & \cite{Charpentier2024}~TC001 & 40 & 3360 & 2.24 & 6.23 & & 1790 & $3.5 \, 10^{-3}$ & 1 & 0.12 \\
a-InO & \cite{Charpentier2024}~TC001 & 40 & 3360 & 2.24 & 3.41 & & 1790 & $6.5 \, 10^{-3}$ & 1 & 0.10 \\
a-InO & DP-res9 & 40 &  &  & 1.66 &  & 2300 & 0.01 & 1 & 0.05\\
a-InO & \cite{Charpentier2024}~TC007-2 & 40 & 5950 & 1.6 & 3.98 & & 4060 & $2.4 \, 10^{-3}$ & 1 & 0.088 \\
a-InO & \cite{Charpentier2024}~TC007-1 & 40 & 7470 & 1.4 & 3.86 & & 5680 & $1.8 \, 10^{-3}$ & 1 & 0.079 \\
a-InO & \cite{Charpentier2024}~TC016-8 & 40 & 12100 & 0.67 & 3.56 & & 11640 & $0.96 \, 10^{-3}$ & 1 & 0.043 \\
a-InO & \cite{Charpentier2024}~TC017 & 40 & 14250 & 0.49 & 3.79 & & 10650 & $0.98 \, 10^{-3}$ & 1 & 0.03 \\
a-InO & \cite{Charpentier2024}~TC016-6 & 40 & 15950 & 0.47 & 2.77 & & 16680 & $0.86 \, 10^{-3}$ & 1 & 0.025 \\
\midrule
Nb & \cite{Verjauw2021} & 100 & 0.293 & 9.02 & 4.137 & 60 &  & 85 & & 32.62 \\
Nb & \cite{Verjauw2021} & 100 & 0.293 & 9.02 & 4.787 & 60 &  & 73 & & 40.22 \\
Nb & \cite{Verjauw2021} & 100 & 0.293 & 9.02 & 5.367 & 60 &  & 65 & & 42.80 \\
Nb & \cite{Verjauw2021} & 100 & 0.293 & 9.02 & 4.503 & 60 &  & 78 & & 47.25 \\
Nb & \cite{Verjauw2021} & 100 & 0.293 & 9.02 & 3.868 & 60 &  & 91 & & 24.31 \\
Nb & \cite{Verjauw2021} & 100 & 0.293 & 9.02 & 5.534 & 60 &  & 63 & & 49.44 \\
Nb & \cite{Zhu2022} & 150 & 0.23 & 9.3 & 6.417 & 64 &  & 48 & 1.2 $10^{-2}$ & 4 \\
Nb & \cite{Altoe2022} & 180 &  &  & 6.5 & 64 &  & 47.5 & & 6 \\
Nb & \cite{Noguchi2019} & 190 & 0.0147 & 9.4 & 4.887 & 32 \cite{Varmazis1974} &  & 253 & 3.8 $10^{-3}$ & 75 \\ 
Nb & \cite{Wang24} & 400 & 0.012 & 9.2 & 5.34 & 38.6 &  & 15.8 &  3.8 $10^{-2}$ & 1.5 \\ 
Nb & \cite{Wang24} & 400 & 0.012 & 9.2 & 5.4 & 38.6 &  & 15.7 &  1.9 $10^{-2}$ & 3.1 \\ 
Nb & \cite{Wang24} & 400 & 0.012 & 9.2 & 5.56 & 38.6 &  & 15.2 &  1.2 $10^{-2}$ & 3.6 \\ 
Nb & \cite{Wang24} & 400 & 0.012 & 9.2 & 5.73 & 38.6 &  & 14.7 &  7.5 $10^{-3}$ & 5.6 \\ 
Nb & \cite{Wang24} & 400 & 0.012 & 9.2 & 5.93 & 38.6 &  & 14.3 &  5.3 $10^{-3}$ & 6.7 \\ 
Nb & \cite{Drimmer24} & 185 & 0.21 &  & 7 & 70.4 &  & 36.5 & 2 $10^{-2}$ & 3.4 \\ 
Nb & \cite{Drimmer24} & 185 & 0.21 &  & 7.5 & 70.4  &  & 34 & 2 $10^{-2}$ & 2.5 \\ 
Nb & \cite{Drimmer24} & 185 & 0.21 &  & 8 & 70.4 &  & 31.9 & 2 $10^{-2}$ & 2.3 \\ 
Nb & \cite{Drimmer24} & 185 & 0.21 &  & 8.5 & 70.4 &  & 30 & 2 $10^{-2}$ & 1.8 \\ 
Nb & \cite{Drimmer24} & 185 & 0.12 &  & 7 &  57.3 &  & 55 & 1.6 $10^{-2}$ & 9.3 \\ 
Nb & \cite{Drimmer24} & 185 & 0.12 &  & 7.5 & 57.3 &  & 51.4 & 1.6 $10^{-2}$ & 7.2 \\ 
Nb & \cite{Drimmer24} & 185 & 0.12 &  & 8 &  57.3 &  & 48.2 & 1.6 $10^{-2}$ & 10.7 \\ 
Nb/Au & \cite{deOry2024} & 112 &  &  & 1.753 &  & 0.24 & 33.5 & 0.1 & 14.8 \\ 
Nb Transmon & \cite{Wang_Transmon_2022} & 120 & 0.22 & 9.1 & Tab. I & 57 & 0.033 &  &  & Tab. I \\ 
Nb Transmon & \cite{Bal_Nb_Transmon_2024} &  &  &  &  SI Tab. I,II & 60* &  &  &  & SI Tab. I,II \\  
Nb Transmon & \cite{Kurter_NbN_Transmon2022} & 200 &  &  & 5* & $\sim$ 100* &  &  &  & Fig. 2b\\
Nb Transmon & \cite{Tuokkola2024} & 200 &  &  & 2.89 & 40 &  & 273 &  & 91\\
Nb Transmon & \cite{Tuokkola2024} & 200 &  &  & 2.89 & 40 &  & 273 &  & 138.9\\

\midrule
Nb thick & \cite{Abdisatarov2024} & 6100 & 1.3 $10^{-3}$ &  & 1.3 & 42.5 &  & 538 & 1.6 $10^{-6}$ & 3 $10^4$\\ 
Nb (3D) & \cite{Romanenko2020} & bulk &  &  & 1.3 & 32 \cite{Varmazis1974} &  & 951 & 1.2 $10^{-6}$ & 2.3 $10^4$ \\ 
Nb (3D) & \cite{Romanenko2020} & bulk &  &  & 1.3 & 32 \cite{Varmazis1974} &  & 951 & 1.2 $10^{-6}$ & 15 $10^4$ \\ 
Nb (3D) & \cite{Romanenko2020} & bulk &  &  & 2.6 & 32 \cite{Varmazis1974} &  & 475 & 2.5 $10^{-6}$ & 1.5 $10^4$ \\ 
Nb (3D) & \cite{Romanenko2020} & bulk &  &  & 5.0 & 32 \cite{Varmazis1974} &  & 247 & 4.8 $10^{-6}$ & 1.0 $10^4$ \\ 
Nb (3D) & \cite{Romanenko2020} & bulk &  &  & 5.0 & 32 \cite{Varmazis1974} &  & 247 & 4.8 $10^{-6}$ & 15 $10^4$ \\ 
Nb (3D) & \cite{Oriani2022} & bulk &  &  & 6.6 & 32 \cite{Varmazis1974} &  & 187 & & 1.5 $10^4$ \\ 
Nb (3D) & \cite{Heidler2021} & bulk &  &  & 7.9 & 39.6 &  & 102 & 3.3 $10^{-5}$ & 0.5 $10^4$ \\
Nb (3D) & \cite{Milul2023} & bulk &  &  & 4.3 & 33 &  & 270 & $5.3 \, 10^{-6}$ & 3 $10^4$\\
Nb (3D) & \cite{Oriani2024} & bulk &  &  & 6.5* & 33 &  & 179 &  & \cite{Oriani2024} Fig. 1 \\
Nb (3D) & \cite{Suleymanzade2020} & bulk &  &  & 98.2 & 40 &  & 8.06 &  & 300\\
Nb (3D) & \cite{Kim2025} & bulk &  &  & 5.779 & 33 &  & 201 &  & 9.4 $10^3$\\
Nb (3D) & \cite{Kim2025} & bulk &  &  & 6.872 & 33 &  & 169 &  & 8.4 $10^3$\\
Nb (3D) & \cite{Esmenda2025} & bulk &  &  & \cite{Esmenda2025} Tab. I & 194 &  &  &  & \cite{Esmenda2025} Tab. I\\

\midrule
NbN & \cite{Frasca2023} & 13 & 106 & 5 & 5.58 & & 34.5 & 0.6 & 1 & 0.861 \\ 
NbN & \cite{Frasca2023} & 13 & 122 & 5.2 & 5.87 & & 38.2 & 0.5 & 1 & 0.753 \\ 
NbN & \cite{Frasca2023} & 13 & 196 & 7.5 & 5.95 & & 44.4 & 0.5 & 1 & 0.885 \\ 
NbN & \cite{Frasca2023} & 13 & 225 & 6.5 & 5.49 & & 52.5 & 0.4 & 1 & 0.59 \\ 
NbN & \cite{Frasca2023} & 13 & 267 & 6 & 5.72 & & 57.2 & 0.35 & 1 & 0.768 \\ 
NbN & \cite{Frasca2023} & 13 & 310 & 5.8 & 4.93 & & 76.8 & 0.3 & 1 & 0.463 \\ 
NbN & \cite{Frasca2023} & 13 & 362 & 5.6 & 6.47 & & 91.3 & 0.2 & 1 & 0.993 \\ 
NbN & \cite{Frasca2023} & 13 & 518 & 4.2 & 3.34 & & 173.3 & 0.2 & 1 & 0.176 \\ 
NbN & \cite{Niepce2019} & 20 & 503 & 7.2 & 4.835 & & 83 & 0.2 & 1 & 0.25 \\ 
NbN & \cite{Foshat2023} & 100 &  &  & 5.952 & & 0.178 & 15 & 0.1 & 0.957 \\ 
NbN & \cite{Yu2021} & 10 & 1033 & 7.4 & 5.7 & & 192 & 0.14 & 0.94 & 0.3 \\ 
NbN & \cite{Yu2021} & 10 & 1033 & 7.4 & 5.756 & & 192 & 0.14 & 1 & 1.58 \\ 
NbN & \cite{Yu2021} & 10 & 1033 & 7.4 & 5.56 & & 192 & 0.15 & 1 & 3.24 \\ 
NbN & \cite{Wei2020} & 6 & 959 & 8.5 & 6.835 & & 132 & 0.3 & 1 & 0.75 \\ 
NbN & \cite{Roy2025} & 13 &  &  & \cite{Roy2025} Tab. II & & 89 &  &  & \cite{Roy2025} Tab. II \\ 
NbN Transmon & \cite{Wei_NbN_Transmon2023} & 8 &  &  & Fig. 7b &  & 200 &  &  & Fig. 7b\\
NbN Transmon & \cite{Kurter_NbN_Transmon2022} & 400 &  &  & 5* & $\sim$ 300* & $\sim$ 0.3* &  &  & Fig. 2b\\
\midrule
NbSi & \cite{LeSueur2018} & 15 & 600 & 0.85 & 6.21 & & 830 & $2.05 \, 10^{-2}$ & & 0.019\\
NbSi & \cite{LeSueur2018} & 15 & 600 & 0.85 & 6.44 & & 830 & $1.98 \, 10^{-2}$ & & 0.023\\
NbSi & \cite{LeSueur2018} & 15 & 600 & 0.85 & 6.73 & & 830 & $1.89 \, 10^{-2}$ & & 0.020\\
NbSi & \cite{LeSueur2018} & 15 & 600 & 0.85 & 7.00 & & 830 & $1.82 \, 10^{-2}$ & & 0.021\\
NbSi & \cite{LeSueur2018} & 15 & 600 & 0.85 & 7.29 & & 830 & $1.75 \, 10^{-2}$ & & 0.026\\ \midrule
NbTiN & \cite{Samkharadze2016} & 8 & 250 (RT) & 9.3 & 2.77 & & 35 & 2 & & 2.11 \\
NbTiN & \cite{Samkharadze2016} & 8 & 250 (RT) & 9.3 & 2.98 & & 35 & 1.9 & & 2.18 \\
NbTiN & \cite{Samkharadze2016} & 8 & 250 (RT) & 9.3 & 3.72 & & 35 & 1.5 & & 1.32 \\
NbTiN & \cite{Samkharadze2016} & 8 & 250 (RT) & 9.3 & 4.06 & & 35 & 1.4 & & 1.46 \\
NbTiN & \cite{Samkharadze2016} & 8 & 250 (RT) & 9.3 & 4.54 & & 35 & 1.25 & & 2.28 \\
NbTiN & \cite{Bruno2015} & 300 & 3.66 (RT?) & 15.5 & 2.75 & & 0.32 & 6 & & 20.9 \\
NbTiN & \cite{Bruno2015} & 300 & 3.66 (RT?) & 15.5 & 3.68 & & 0.32 & 4.5 & & 18.3 \\
NbTiN & \cite{Bruno2015} & 300 & 3.66 (RT?) & 15.5 & 4.57 & & 0.32 & 3.6 & & 13.5 \\
NbTiN & \cite{Bruno2015} & 300 & 3.66 (RT?) & 15.5 & 6.41 & & 0.32 & 2.6 & & 20.2 \\ 
NbTiN & \cite{Giachero2023} & 5 & 817 & 9.5 & 2.5* & & 118.6 & 1 & & 0.3 \\
NbTiN & \cite{Giachero2023} & 10 & 277 & 11.6 & 2* & & 32.9 & 2.4 & & 0.5 \\
NbTiN & \cite{Giachero2023} & 20 & 100 & 13.1 & 3* & & 10.5 & 2.5 & & 0.7 \\
NbTiN & \cite{Yang24} & 12 &  &  & 7.03 & & 1248 & 0.015 & & 0.14 \\
NbTiN & \cite{Yang24} & 12 &  &  & 8.08 & & 965 & 0.017 & & 0.11 \\
NbTiN & \cite{Yang24} & 12 &  &  & 7.02 & & 1252 & 0.015 & & 0.21 \\
NbTiN & \cite{Cools2025} & 10 &  &  & \cite{Cools2025} Tab. I & & \cite{Cools2025} Tab. I &  & & \cite{Cools2025} Tab. I \\
\midrule
PtSi & \cite{Szypryt2016} & 60 &  &  & 4.876 & & 8.2 & 0.66 & & 1.47 \\ 
\midrule
Re & \cite{Dumur2016} & 25 & 1.905 & 1.85 & 6.034 & & 1.42 & 7.5 & & 0.14 \\
Re & \cite{Dumur2016} & 100 & 0.054 & 1.88 & 6.99 & & 0.04 & 57 & & 0.08 \\ 
Re & \cite{Wang2026} & 150 & 0.035 & 1.9 & \cite{Wang2026} Table I & $\sim 150$ &  & 11.2 & & \cite{Wang2026} Table I \\ 
\midrule
$\alpha$-Ta & \cite{Crowley23} & 200 &  &  & 4.484 & & 0.063* & 28 & & 95 \\
$\alpha$-Ta & \cite{Lozano2022} & 100 & 1.0 & 2.9 & 6* & & 0.476 & 5.5 & & 5.5 \\
$\alpha$-Ta & \cite{Lozano2022} & 100 & 1.0 & 2.9 & 6* & & 0.476 & 5.5 & & 11.5 \\
$\alpha$-Ta & \cite{Lozano2022} & 100 & 1.0 & 2.9 & 6* & & 0.476 & 5.5 & & 7 \\
$\alpha$-Ta & \cite{Lozano2022} & 100 & 1.0 & 2.9 & 6* & & 0.476 & 5.5 & & 8.8 \\
$\alpha$-Ta & \cite{Lozano2022} & 100 & 1.0 & 2.9 & 6* & & 0.476 & 5.5 & & 12.8 \\
$\alpha$-Ta & \cite{Lozano2022} & 100 & 1.0 & 2.9 & 6* & & 0.476 & 5.5 & & 25.1 \\
$\alpha$-Ta & \cite{Lozano2022} & 100 & 1.0 & 2.9 & 6* & & 0.476 & 5.5 & & 10.6 \\
$\alpha$-Ta & \cite{Lozano2022} & 100 & 1.0 & 2.9 & 6* & & 0.476 & 5.5 & & 33.3 \\
$\alpha$-Ta & \cite{Lozano2022} & 100 & 1.0 & 2.9 & 6* & & 0.476 & 5.5 & & 23.7 \\
$\alpha$-Ta & \cite{Jia2023} & 150 & 0.1* & 4.12 & 5.1 & & 0.033* & 62 & $7.8 \, 10^{-3}$ & 14 \\
$\alpha$-Ta & \cite{Jia2023} & 150 & 0.1* & 4.12 & 5.4 & & 0.033* & 58.5 & $7.8 \, 10^{-3}$ & 14 \\
$\alpha$-Ta & \cite{Jia2023} & 150 & 0.1* & 4.12 & 5.9 & & 0.033* & 53.5 & $7.8 \, 10^{-3}$ & 12.5 \\
$\alpha$-Ta & \cite{Jia2023} & 150 & 0.1* & 4.12 & 6.2 & & 0.033* & 51 & $7.8 \, 10^{-3}$ & 12 \\
$\alpha$-Ta & \cite{Jia2023} & 150 & 0.1* & 4.12 & 6.4 & & 0.033* & 49.5 & $7.8 \, 10^{-3}$ & 15 \\
$\alpha$-Ta & \cite{Jia2023} & 150 & 0.1* & 4.12 & 6.8 & & 0.033* & 46.5 & $7.8 \, 10^{-3}$ & 10 \\
$\alpha$-Ta & \cite{Jia2023} & 150 & 0.1* & 4.12 & 7.1 & & 0.033* & 44.5 & $7.8 \, 10^{-3}$ & 7.5 \\
$\alpha$-Ta & \cite{Shi2022} & 200 & 0.1 & 4.42 & Fig. 3b & & 0.031 & 24.5 -- 29 & $7.7 \, 10^{-3}$ & Fig. 3b \\
$\alpha$-Ta & \cite{Schijndel2024} & 50 & 0.15 & 4.14 & 5.5 & 45 & 0.071 & 81.3 & $2.8 \, 10^{-2}$ & 19 \\
$\alpha$-Ta & \cite{Schijndel2024} & 50 & 0.15 & 4.14 & 6.5 & 45 & 0.071 & 68.8 & $2.8 \, 10^{-2}$ & 7 \\
$\alpha$-Ta & \cite{Li2024} & 300 &  &  & 6.63 & 150 & 0.195 & 8.48 & $2.4 \, 10^{-2}$ & 25 \\
$\alpha$-Ta & \cite{Marcaud2025} & 100 & 0.078 & 4.3 & 6 &  & 0.025 & 106 &  & 41.6 \\
$\alpha$-Ta & \cite{Marcaud2025} & 100 & 0.078 & 4.3 & 5.3 &  & 0.025 & 120 &  & 34.4 \\
$\alpha$-Ta & \cite{Marcaud2025} & 100 & 1.16 & 4.45 & 5.4 &  & 0.36 & 8.2 &  & 62.5 \\
$\alpha$-Ta & \cite{Marcaud2025} & 100 & 1.16 & 4.45 & 4.6 &  & 0.36 & 9.6 &  & 24.3 \\
$\alpha$-Ta & \cite{Ganjam24} & 150 & 0.015 & 4.3 & \cite{Ganjam24} Tab. S5 & 150 &  & &  &  \cite{Ganjam24} Tab. S1\\
$\alpha$-Ta & \cite{Zikiy2025} & 150 & 1.2 & 4.39 & 4.8 & 212 &  & 5.85 &  & 91.5 \\
$\alpha$-Ta Transmon & \cite{Wang_Transmon_2022} & 120 & 0.24 & 4.2 & Tab. I & & 0.08 &  &  &  Tab. I \\ 
$\alpha$-Ta Transmon & \cite{Place_Ta_Transmon_2021} & 200 & 0.2 & 4.38 &  SI Tab. I & & 0.063 &  &  & SI Tab. I \\ 
$\alpha$-Ta Transmon & \cite{Bland2025} & 200* & 0.27* & 4.2 &  \cite{Bland2025} Tab. I & 150* & &  &  & \cite{Bland2025} Tab. I\\ 
$\beta$-Ta & \cite{Kouwenhoven2023} & 60 & 39.8 & 1.0 & 8* & & 54.6 & 0.060 & & 4.3 \\ 
$\beta$-Ta & \cite{Joshi2026} &  &  & 0.7 &  & 1780 &  & \cite{Joshi2026} Fig. S7 & & \cite{Joshi2026} Fig. S7 \\ 
$\beta$-Ta Transmon & \cite{Joshi2026} &  &  & 0.7 & 2.74 & 1780 &  & 0.145 & & 56 \\
$\beta$-Ta Transmon & \cite{Joshi2026} &  &  & 0.7 & 5.80 & 1780 &  & 0.07 & & 15 \\ \midrule
TiAlV (3D) & \cite{Holland2017} & bulk &  & 4.5 & 7.5 & 8000 & & $2.63 \, 10^{-3}$ & $4 \, 10^{-3}$ & 0.3\\ 
TiAlV (3D) & \cite{Carriere2022} & bulk &  & 3.5 & 5.917 & 657 & 0.82 & 0.49 &  & 22\\ \midrule
TiN & \cite{Amin2022} & 3 & 522 & 3 & 5.563 & & 239 & 0.4 & 0.99 & 0.9 \\ 
TiN & \cite{Richardson2020} & 60 & 3.23 & 5.4 & 5.5* & & 0.82 & 5.9 & 0.25 & 0.95 \\
TiN & \cite{Richardson2020} & 60 & 3.23 & 5.4 & 5.5* & & 0.82 & 5.9 & 0.25 & 0.72 \\
TiN & \cite{Richardson2020} & 60 & 3.23 & 5.4 & 5.5* & & 0.82 & 5.9 & 0.25 & 0.61 \\
TiN & \cite{Richardson2020} & 60 & 3.23 & 5.4 & 5.5* & & 0.82 & 5.9 & 0.25 & 0.54 \\
TiN & \cite{Richardson2020} & 60 & 3.23 & 5.4 & 5.5* & & 0.82 & 5.9 & 0.25 & 0.51 \\
TiN & \cite{Gao2022} & 100 & 9 & 3.8 & 6* & & 3.27 & 0.8 & 0.43 & 3.3 \\
TiN & \cite{Gao2022} & 100 & 9 & 3.8 & 6* & & 3.27 & 0.8 & 0.43 & 1.8 \\
TiN & \cite{Shearrow2018} & 8.9 & 505 (RT) & 3.01 & 5* & & 234 & 0.15 & & \cite{Shearrow2018} Fig.2c \\
TiN & \cite{Shearrow2018} & 14.2 & 145 (RT) & 3.63 & 5* & & 56 & 0.4 & & \cite{Shearrow2018} Fig.2c \\
TiN & \cite{Shearrow2018} & 49.8 & 20.7 (RT) & 4.05 & 5* & & 7.1 & 0.9 & & \cite{Shearrow2018} Fig.2c \\
TiN & \cite{Shearrow2018} & 109 & 5.68 (RT) & 4.62 & 5* & & 1.7 & 1.7 & & \cite{Shearrow2018} Fig.2c \\ 
TiN & \cite{Mencia2021} & 100 & 316 & 2.95 & 10* & & 148 & 0.01 & & 0.022 \\ 
TiN & \cite{Ohya2013} & 100 & 10 & 3.5 & 5* & & 3.94 & 0.80 & & 50 \\ 
TiN & \cite{Koolstra25} & 12 & 361 & 2.8 & 5.025 & & 178 & 0.15 & & 39 \\ 
TiN & \cite{Wu2024} & 35 &  &  & \cite{Wu2024} Fig. 5& & \cite{Wu2024} Tab. 2 &  & & \cite{Wu2024} Fig. 5\\
TiN & \cite{Li2025a} & 27 &  &  & 0.92 & & 394 & 0.162 & & 0.41\\
TiN & \cite{Li2025a} & 27 &  &  & 1.00 & & 394 & 0.149 & & 0.39\\
TiN & \cite{Li2025a} & 27 &  &  & 1.10 & & 394 & 0.136 & & 0.43\\
TiN & \cite{Tominaga2025} & 100 & 0.25 & 5.6 & 4.49 & & 0.061 & 57.5 & & 17.3\\
TiN & \cite{Tominaga2025} & 100 & 0.25 & 5.6 & 5.34 & & 0.061 & 48.3 & & 25.8\\
TiN & \cite{Tominaga2025} & 100 & 0.25 & 5.6 & 6.17 & & 0.061 & 41.8 & & 8.13\\
TiN & \cite{Tominaga2025} & 100 & 0.25 & 5.6 & 4.02 & & 0.061 & 64.2 & & 63\\
TiN & \cite{Tominaga2025} & 100 & 0.25 & 5.6 & 4.81 & & 0.061 & 53.7 & & 45\\
TiN & \cite{Tominaga2025} & 100 & 0.25 & 5.6 & 6.00 & & 0.061 & 43.0 & & 96\\
TiN & \cite{Tominaga2025} & 100 & 0.25 & 5.6 & 6.85 & & 0.061 & 37.7 & & 84.0\\
TiN Transmon & \cite{Deng_Transmon_2023} & 100 & 9 & 3.8 & Tab. I &  & 3.27~\cite{Gao2022}&  &  & Tab. I\\
\midrule
TiN/Al & \cite{Gao2022b} & 100 & 0.1323 & 3.2 & 4.3733 & & 57 & 0.06 & 0.88 & 2.7 \\
TiN/Al & \cite{Gao2024} & 10 &  &  & 4.300 & & 1200 & $3.08 \, 10^{-2}$ & 1 & 4.54 \\
TiN/Al & \cite{Gao2024} & 10 &  &  & 4.370 & & 1200 & $3.03 \, 10^{-2}$ & 1 & 4.47 \\
TiN/Al & \cite{Gao2024} & 10 &  &  & 4.407 & & 1200 & $3.00 \, 10^{-2}$ & 1 & 3.61 \\
TiN/Al & \cite{Gao2024} & 10 &  &  & 4.418 & & 1200 & $3.00 \, 10^{-2}$ & 1 & 2.61 \\
TiN/Al & \cite{Gao2024} & 10 &  &  & 4.500 & & 1200 & $2.94 \, 10^{-2}$ & 1 & 4.11 \\
TiN/Al & \cite{Gao2024} & 10 &  &  & 4.537 & & 1200 & $2.92 \, 10^{-2}$ & 1 & 3.07 \\
TiN/Al & \cite{Gao2024} & 10 &  &  & 4.547 & & 1200 & $2.91 \, 10^{-2}$ & 1 & 2.96 \\
TiN/Al & \cite{Gao2024} & 20 &  &  & 7.298 & & 6500 & $1.67 \, 10^{-3}$ & 1 & 0.187 \\
TiN/Al & \cite{Gao2024} & 20 &  &  & 7.529 & & 6500 & $1.62 \, 10^{-3}$ & 1 & 0.145 \\
TiN/Al & \cite{Gao2024} & 20 &  &  & 7.697 & & 6500 & $1.59 \, 10^{-3}$ & 1 & 0.160 \\
TiN/Al & \cite{Gao2024} & 20 &  &  & 7.819 & & 6500 & $1.56 \, 10^{-3}$ & 1 & 0.155 \\
TiN/Al & \cite{Gao2024} & 20 &  &  & 7.896 & & 6500 & $1.55 \, 10^{-3}$ & 1 & 0.156 \\
TiN/Al & \cite{Gao2024} & 20 &  &  & 7.948 & & 6500 & $1.54 \, 10^{-3}$ & 1 & 0.177 \\
TiN/Al & \cite{Gao2024} & 20 &  &  & 8.064 & & 6500 & $1.51 \, 10^{-3}$ & 1 & 0.161 \\ 
\midrule
WSi & \cite{Quaranta2013} & 100 & 20 & 4 & 1.766 & & 6.9 & 1.3 & 0.92 & 7.1 \\ 
WSi & \cite{Quaranta2013} & 365 & 12.32 & 1.8 & 4.705 & & 9.46 & 0.09 & 0.42 & 1.7 \\ 
WSi & \cite{Kirsh2021} & 30 & 98 & 4.75 & 1.032 & & 30 & 1.7 & 0.88 & 1.4 \\ 
WSi & \cite{Larson2025b} & 3 & &  & \cite{Larson2025b}Fig. 1d & & 300 &  &  & \cite{Larson2025b}Fig. 1d \\ 
WSi & \cite{Larson2025b} & 10 & &  & \cite{Larson2025b}Fig. 1d & & 100 &  &  & \cite{Larson2025b}Fig. 1d \\ 
\midrule
\end{longtable*}
}

\end{document}